# Stress Tensor Eigenvector Following with Next-Generation Quantum Theory of Atoms in Molecules

Jia Hui Li, Wei Jie Huang, Tianlv Xu, Steven R. Kirk[*] and Samantha Jenkins[*]

*Key Laboratory of Chemical Biology and Traditional Chinese Medicine Research and Key Laboratory of Resource Fine-Processing and Advanced Materials of Hunan Province of MOE, College of Chemistry and Chemical Engineering, Hunan Normal University, Changsha, Hunan 410081, China*

email: steven.kirk@cantab.net
email: samanthajsuman@gmail.com

The eigenvectors of the electronic stress tensor have been identified as useful for the prediction of chemical reactivity because they determine the most preferred directions to move the bonds that correspond to a qualitative change in the molecular electronic structure. A new 3-D vector based interpretation of the chemical bond that we refer to as the bond-path framework set $\mathbb{B}$ = {***p***,***q***,***r***} provides a version of the quantum theory of atoms in molecules (QTAIM) beyond the minimum definition for bonding that is particularly suitable for understanding changes in molecular electronic structure that occur during reactions. The bond-path framework set $\mathbb{B}$ is straightforwardly constructed and visualized from the eigenvalues and eigenvectors of QTAIM. This approach is applied to the structural deformations of ethene that occur during applied torsion θ, -180.0° ≤ θ ≤ +180.0°. The corresponding stress tensor version is readily constructed as $\mathbb{B}_\sigma$ = {***p**$_\sigma$*,***q**$_\sigma$*,***r***} within the QTAIM partitioning making it possible to compare experimentally and computationally determined electronic charge densities. The bond-path framework set $\mathbb{B}$ or $\mathbb{B}_\sigma$ are the networks that comprise three strands: the least preferred (***p***,***p**$_\sigma$*), most preferred (***q***,***q**$_\sigma$*) and ***r*** is the familiar QTAIM bond-path. We demonstrate that the most preferred direction for bond motion using the stress tensor corresponds to the most compressible direction and not to the least compressible direction as previously reported. We show the necessity for a directional approach constructed using the eigenvectors along the entire bond length and demonstrate the insufficiency of the sole use of scalar measures for capturing the nature of the stress tensor within the QTAIM partitioning.

## Introduction

The quantum stress tensor, σ(**r**) is directly related to the Ehrenfest force by the virial theorem and therefore provides a physical explanation of the low frequency normal modes that accompany structural rearrangements[1]. In this work we use the definition of the stress tensor proposed by Bader[2] to investigate the stress tensor properties within the quantum theory of atoms in molecules (QTAIM) partitioning scheme[3]. Earlier, it was found that the stress tensor properties such as the stiffness, $\mathbb{S}_\sigma = |\lambda_{1\sigma}|/|\lambda_{3\sigma}|$ produced results that were in line with physical intuition[1,4] as well as the stress tensor trajectories $\mathbb{T}_\sigma(s)$[5].

If we first consider a tiny cube of fluid flowing in 3-D space the stress $\Pi(x, y, z, t)$, a rank-3 tensor field, has nine components[6] of these the three diagonal components $\Pi_{xx}$, $\Pi_{yy}$, and $\Pi_{zz}$ correspond to normal stress. A negative positive value for these normal components signifies a compression of the cube, conversely a positive value refers to pulling or tension, where more negative/positive values correspond to increased compression/tension of the cube. Diagonalization of the stress tensor, σ(**r**), returns the principal electronic stresses $\Pi_{xx}$, $\Pi_{yy}$, and $\Pi_{zz}$ that are realized as the stress tensor eigenvalues $\lambda_{1\sigma}$, $\lambda_{2\sigma}$, $\lambda_{3\sigma}$, with corresponding eigenvectors **e**$_{1\sigma}$, **e**$_{2\sigma}$, **e**$_{3\sigma}$ are calculated within the QTAIM partitioning. The interpretation of the eigenvalues is different between QTAIM and the stress tensor: in QTAIM the most 'easy' preferred direction is simply the shallowest direction based on the readiness of the electronic charge density to accumulate or move. For the stress tensor however, the most preferred 'easy' direction is determined as the most compressible, i.e. the least tensile. The eigenvalues are ordered $\lambda_{1\sigma} < \lambda_{2\sigma} < \lambda_{3\sigma}$ for the stress tensor with $\lambda_{3\sigma}$ being the purely tensile and $\lambda_{1\sigma}$ being the most compressive. For QTAIM the ordering is $\lambda_1 < \lambda_2 < \lambda_3$ with $\lambda_2$ being shallower and more changeable than $\lambda_1$, enabling us to understand that $\lambda_2$ is comparable most compressible $\lambda_{1\sigma}$ stress tensor eigenvalue. Consequently, the stress tensor eigenvectors **e**$_{1\sigma}$ and **e**$_{2\sigma}$ frequently do not coincide with the QTAIM **e**$_1$ and **e**$_2$ eigenvectors respectively, particularly for symmetrical bonds such as the central C-C bond in biphenyl that links the two phenyl rings.

The purpose of this investigation is to determine how to use the stress tensor within the QTAIM partitioning as originally envisaged from the observations of the normal phonon modes of ice and the QTAIM eigenvectors[7]. A central theme of this work therefore will be to develop an in depth understanding of the *directional* character of the stress tensor i.e. the **e**$_{1\sigma}$ and **e**$_{2\sigma}$ eigenvectors as opposed to only considering the scalar eigenvalues $\lambda_{1\sigma}$ and $\lambda_{2\sigma}$, perhaps in the form of a stress tensor ellipticity $\varepsilon_\sigma = |\lambda_{2\sigma}|/|\lambda_{1\sigma}| - 1$ or $\varepsilon_{\sigma H} = |\lambda_{1\sigma}|/|\lambda_{2\sigma}| - 1$, where the subscript 'H' is used to denote the Hessian numerator and denominator ordering. The earlier attempt assumed that there was a one-to-one mapping between the QTAIM least preferred **e**$_1$ and most preferred **e**$_2$ directions of motions of the electronic charge density and the stress tensor **e**$_{1\sigma}$ and **e**$_{2\sigma}$ eigenvectors[8] or even only the eigenvalues[9]. Later investigations have demonstrated a lack of one to one mapping between the eigenvectors and eigenvalues of QTAIM and the stress tensor mapping[5,10]. This confusion arose due to the accidental coincidence of the directions of the QTAIM **e**$_2$ and stress tensor **e**$_{2\sigma}$ eigenvectors for bond-paths with *BCP*s located away from the geometric mid-point that occurs for asymmetrical bonds. Such results were misleading and arose as a consequence of the use of the stress tensor within the QTAIM partitioning where for asymmetrically located *BCP*s such as occur for the bond-path of the C-H *BCP* the $-\nabla\cdot\sigma(\mathbf{r_b}) \neq 0$ unlike the QTAIM result $\nabla\cdot\rho(\mathbf{r_b}) = 0$. In this investigation therefore, we consider the QTAIM and stress tensor eigenvectors along the *entire* bond-path rather than only at the *BCP*, to avoid such misleading results.

A further consequence of this mismatch in the positions of $\nabla\cdot\rho(\mathbf{r_b}) = 0$ and $-\nabla\cdot\sigma(\mathbf{r_b}) = 0$ will be that the stress tensor properties derived from the stress tensor eigenvalues will be sensitive to small variations, where previously we actually employed $\lambda_{3\sigma} < 0$ as a useful measure of instability or approaching phase transition[1,11]. The goal of this work is to understand how to more reliably use the stress tensor results within the QTAIM partitioning scheme. A part of this work will involve the creation of a non-minimal interpretation of the chemical bond, comprising three strands or *paths*, that can be rendered in 3-D to visualize the most and least preferred directions of bond motion in addition to the familiar and minimal bond-path. This will be undertaken by attempting to understand the relationship between the compressive stress tensor **e**$_{1\sigma}$, **e**$_{2\sigma}$ and QTAIM **e**$_1$, **e**$_2$ eigenvectors.

**2. Theory and Methods**
*2.1 The QTAIM and stress tensor BCP properties; the ellipticity ε and the stress ellipticities $\varepsilon_{\sigma H}$ and $\varepsilon_\sigma$*

We use QTAIM and the stress tensor analysis that utilizes higher derivatives of $\rho(\mathbf{r_b})$ in effect, acting as a 'magnifying lens' on the $\rho(\mathbf{r_b})$ derived properties of the wave-function. Current representations of the chemical bond, also within the QTAIM framework include bond-bundles in open systems, whereby molecules are partitioned through an extension of QTAIM where bounded regions of space containing non-bonding or lone-pair electrons are created that lead to bond orders consistent with expectation from theories of directed valence[12,13]. We will use QTAIM[3] to identify critical points in the total electronic charge density distribution $\rho(\mathbf{r})$ by analyzing the gradient vector field $\nabla\rho(\mathbf{r})$. These critical points can further be divided into four types of topologically stable critical points according to the set of ordered eigenvalues $\lambda_1 < \lambda_2 < \lambda_3$, with corresponding eigenvectors **e**$_1$, **e**$_2$, **e**$_3$ of the Hessian matrix. The Hessian of the total electronic charge density $\rho(\mathbf{r})$ is defined as the matrix of partial second derivatives with respect to the spatial coordinates. These critical points are labeled using the notation (*R*, ω) where *R* is the rank of the Hessian matrix, the number of distinct non-zero eigenvalues and ω is the signature (the algebraic sum of the signs of the eigenvalues); the (3, -3) [nuclear critical point (*NCP*), a local maximum generally corresponding to a nuclear location], (3, -1) and (3, 1) [saddle points, called bond critical points (*BCP*) and ring critical points (*RCP*), respectively] and (3, 3) [the cage critical points (*CCP*)]. In the limit that the forces on the nuclei become vanishingly small, an atomic interaction line[14] becomes a bond-path, although not necessarily a chemical bond[15]. The complete set of critical points together with the bond-paths of a molecule or cluster is referred to as the molecular graph.

The eigenvector **e**$_3$ indicates the direction of the bond-path at the *BCP*. The most and least preferred directions of electron accumulation are $\underline{\mathbf{e}_2}$ and $\underline{\mathbf{e}_1}$, respectively[16–18]. The ellipticity, ε provides the relative accumulation of $\rho(\mathbf{r_b})$ in the two directions perpendicular to the bond-path at a *BCP*, defined as $\varepsilon = |\lambda_1|/|\lambda_2| - 1$ where $\lambda_1$ and $\lambda_2$ are negative eigenvalues of the corresponding eigenvectors $\underline{\mathbf{e}_1}$ and $\underline{\mathbf{e}_2}$ respectively. Recently, for the 11-cis retinal subjected to a torsion ±θ, we have recently demonstrated that the $\underline{\mathbf{e}_2}$ eigenvector of the torsional *BCP* corresponded to the preferred +θ direction of rotation as defined by the PES profile[19].

In this investigation we will define two ellipticities for the stress tensor:

$$\varepsilon_{\sigma H} = |\lambda_{1\sigma}|/|\lambda_{2\sigma}| - 1 \qquad (1a)$$

$$\varepsilon_{\sigma} = |\lambda_{2\sigma}|/|\lambda_{1\sigma}| - 1 \qquad (1b)$$

Where the subscript 'H' of $\varepsilon_{\sigma H}$ in equation **(1a)** refers to the use of the most/least negative eigenvalues for the eigenvalues of numerator/denominator as the QTAIM ellipticity ε, consequently the stress tensor ellipticity $\varepsilon_{\sigma H} \geq 0$ and the ellipticity $\varepsilon \geq 0$ without exception, due to the eigenvalues being ordered $\lambda_{1\sigma} < \lambda_{2\sigma} < \lambda_{3\sigma}$ and $\lambda_1 < \lambda_2 < \lambda_3$. Conversely, equation **(1b)** that defines $\varepsilon_\sigma$ uses the least/most negative eigenvalues for the eigenvalues of numerator/denominator and due to the eigenvalues being ordered $\lambda_{1\sigma} < \lambda_{2\sigma} < \lambda_{3\sigma}$ then $\varepsilon_\sigma \leq 0$ without exception. The reason we choose the counterintuitive result that $\varepsilon_\sigma \leq 0$, see equation **(1b)** is because for the stress tensor the 'easy' direction ($\underline{\mathbf{e}_{1\sigma}}$) is determined by the most compressible eigenvalue $\lambda_{1\sigma}$ i.e. associated with the longer axis of the ellipse associated with ellipticity. Conversely, for QTAIM the 'easy' direction ($\underline{\mathbf{e}_2}$) is associated with the longer axis ($\lambda_2$) of the ellipse. This is because for QTAIM there is an ellipse shaped distribution in $\rho(\mathbf{r_b})$ for values of $\varepsilon > 0$, perpendicular to the bond-path with long (associated with the 'easy' direction $\underline{\mathbf{e}_2}$) and short (associated with the 'hard' direction $\underline{\mathbf{e}_1}$) axes defined by the $\lambda_2$ and $\lambda_1$ eigenvalues respectively.

*2.2 The QTAIM, $\mathbb{B} = \{p,q,r\}$ and stress tensor bond-path framework set $\mathbb{B}_{\sigma H} = \{p_{\sigma H}, q_{\sigma H}, r\}$ and $\mathbb{B}_\sigma = \{p_\sigma, q_\sigma, r\}$*

The bond-path length (BPL) is defined as the length of the path traced out by the $\underline{\mathbf{e}_3}$ eigenvector of the Hessian of the total charge density $\rho(\mathbf{r})$, passing through the *BCP*, along which $\rho(\mathbf{r})$ is locally maximal with respect to any neighboring paths. The bond-path curvature separating two bonded nuclei is defined as the dimensionless ratio:

$$(BPL - GBL)/GBL, \qquad (2)$$

Where the BPL is the associated bond-path length and the geometric bond length GBL is the inter-nuclear separation. The BPL often exceeds the GBL particularly for weak or strained bonds and unusual bonding environments[20]. Earlier, one of the current authors hypothesized that the morphology of a bond-path may be 1-D i.e. a linear bond-path equal in length to the bonded inter-nuclear separation, bent with one radius of curvature (2-D) only in the direction of $\underline{\mathbf{e}_2}$. For 3-D bond-paths, there are minor and major radii of curvature specified by the directions of **e₂** and **e₁** respectively[21]. In this investigation we suggest the involvement of the **e₃** eigenvector also, in the form of a bond-path twist. It was observed during calculations of the **e₁** and **e₂** eigenvectors at successive points along the bond-path that in some cases, these eigenvectors, both being perpendicular to the bond-path tracing eigenvector **e₃**, 'switched places'. We recently observed that the calculation of the vector tip path following the unscaled **e₁** eigenvector would then show a large 'jump' as it swapped directions with the corresponding **e₂** eigenvector[22]. This phenomenon indicated a location where the ellipticity ε must be zero due to degeneracies in the corresponding $\lambda_1$ and $\lambda_2$ eigenvalues. The choice of the ellipticity ε as scaling factor was motivated by the fact that the scaled vector tip paths drop smoothly onto the bond-path, ensuring that the tip paths are always continuous. We previously discussed the unsuitability of alternative scaling factors, $|\lambda_1 - \lambda_2|$ this was not pursued as it lacks the universal chemical interpretation of the ellipticity ε e.g. double-bond ε > 0.25 vs. single bond character ε ≈ 0.10. Also unsuitable choices for scaling factors, on the basis of not attaining zero, included either ratios involving the $\lambda_1$ and $\lambda_2$ eigenvalue or any inclusion of the $\lambda_3$ eigenvalue. The $\lambda_3$ eigenvalue was also found to unsuitable because it contains no information about the least (**e₁**) and most (**e₂**) preferred directions of the total charge density ρ(**r**) accumulation.

With *n* scaled eigenvector **e₂** tip path points $q_i = r_i + \varepsilon_i \mathbf{e}_{2,i}$ on the *q*-path where $\varepsilon_i$ = ellipticity at the $i^{th}$ bond-path point $r_i$ on the bond-path *r*. It should be noted that the bond-path is associated with the $\lambda_3$ eigenvalues of the **e₃** eigenvector does not take into account differences in the $\lambda_1$ and $\lambda_2$ eigenvalues of the **e₁** and **e₂** eigenvectors. Analogously, for the **e₁** tip path points we have $p_i = r_i + \varepsilon_i \mathbf{e}_{1,i}$ on the *p*-path where $\varepsilon_i$ = ellipticity at the $i^{th}$ bond-path point $r_i$ on the bond-path *r*.

We referred to the next-generation QTAIM interpretation of the chemical bond as the *bond-path framework set*, denoted by $\mathbb{B}$, where $\mathbb{B} = \{p,q,r\}$ with the consequence that for the ground state a bond is comprised of three 'linkages'; *p*-, *q*- and *r*-paths associated with the **e₁**, **e₂** and **e₃** eigenvectors, respectively.

The *p* and *q* parameters define eigenvector-following paths with lengths $\mathbb{H}^*$ and $\mathbb{H}$, see **Scheme 2**:

$$\mathbb{H}^* = \sum_{i=1}^{n-1} |p_{i+1} - p_i| \qquad (3a)$$
$$\mathbb{H} = \sum_{i=1}^{n-1} |q_{i+1} - q_i| \qquad (3b)$$

The lengths of the *eigenvector-following paths* $\mathbb{H}^*$ or $\mathbb{H}$ refers to the fact that the tips of the scaled **e₁** or **e₂** eigenvectors sweep out along the extent of the bond-path, defined by the **e₃** eigenvector, between two bonded nuclei connected by a bond-path. In the limit of vanishing ellipticity ε = 0, *for all* steps *i* along the bond-path then $\mathbb{H}$ = BPL.

From $p_i = r_i + \varepsilon_i \mathbf{e}_{1,i}$ and $q_i = r_i + \varepsilon_i \mathbf{e}_{2,i}$ we see for shared-shell *BCP*s, in the limit of the ellipticity ε ≈ 0 i.e. corresponding to single bonds, we then have $p_i = q_i = r_i$ and therefore the value of the lengths $\mathbb{H}^*$ and $\mathbb{H}$ attain their lowest limit; the bond-path length (*r*) BPL. Conversely, higher values of the ellipticity ε, for instance, corresponding to double bonds will always result in values of $\mathbb{H}^*$ and $\mathbb{H}$ > BPL.

In addition, because $\mathbb{H}^*$ and $\mathbb{H}$ are defined by the distances swept out by the $\underline{e}_2$ tip path points $p_i = r_i + \varepsilon_i \underline{e}_{1,i}$ and $q_i = r_i + \varepsilon_i \underline{e}_{2,i}$ respectively and the scaling factor, $\varepsilon_i$ is identical in equation **(3a)** and equation **(3b)** therefore for a linear bond-path $r$ then $\mathbb{H}^* = \mathbb{H}$. The bond-path framework set $\mathbb{B} = \{p,q,r\}$ should consider the bond-path to comprise the *unique p-, q-* and *r*-paths, swept out by the $\underline{e}_1$, $\underline{e}_2$ and $\underline{e}_3$, eigenvectors that form the eigenvector-following paths with lengths $\mathbb{H}^*$, $\mathbb{H}$ and BPL respectively. The *p-* and *q*-paths are unique even when the lengths of $\mathbb{H}^*$ and $\mathbb{H}$ are the same or very similar because the *p-* and *q*-paths traverse different regions of space. Bond-paths *r* with non-zero bond-path curvature which will result in $\mathbb{H}^*$ and $\mathbb{H}$ with different values, this is more likely to occur for the equilibrium geometries of closed-shell *BCP*s than for shared-shell *BCP*s. This is because the *p-* and *q*-paths will be different because of the greater distance travelled around the outside of a twisted bond-path *r* compared with the inside of the same twisted bond-path *r*. This is because within QTAIM the $\underline{e}_1$, $\underline{e}_2$ and $\underline{e}_3$, eigenvectors can only be defined to within to a factor of -1, i.e. ($\underline{e}_1,-\underline{e}_1$), ($\underline{e}_2,-\underline{e}_2$) and ($\underline{e}_3,-\underline{e}_3$) therefore there will be two possible tip-paths. The consequences of this (within QTAIM) calculation of the $\mathbb{H}^*$ is that we dynamically update the sign convention to define $\mathbb{H}^*$ as being the shorter of the two possible tip-paths because $\underline{e}_1$ is the least preferred direction of accumulation of $\rho(\mathbf{r})$. A similar procedure is used for $\mathbb{H}$ except that we chose the longer of the two possible tip-paths because $\underline{e}_2$ is the most preferred direction of accumulation of $\rho(\mathbf{r})$.

When using the stress tensor to generate $p_\sigma$- and $q_\sigma$-paths, remembering that $\varepsilon_\sigma \leq 0$, particularly for bond-paths *r* where the *BCP is* located away from the geometric mid-point of the bonded nuclei, we note that $p_\sigma$- or $q_\sigma$-paths that lie in the plane of curvature of the bond-path *r*, display a greater variation with applied torsion θ than paths that deviate from this plane. The $p_\sigma$- or $q_\sigma$-paths that lie in the bond-path curvature *r* plane will be expected to reflect the inherent asymmetry of the variation of the $\mathbb{H}^*_\sigma$ or $\mathbb{H}_\sigma$ with the applied torsion θ. This asymmetry exists as a consequence of the $-\nabla \cdot \sigma(\mathbf{r_b})$ possessing non-zero magnitude caused by obtaining all of the stress tensor properties within the QTAIM partitioning. The remaining $p_\sigma$- or $q_\sigma$-path that does not lie in the plane of curvature of the bond-path *r* would be expected to possess $\mathbb{H}^*_\sigma$ or $\mathbb{H}_\sigma$ values that vary symmetrically with an applied torsion θ. In this investigation the C-H bond-paths would be expected to display this asymmetrical and symmetrical character with the applied torsion $-180.0° \leq θ \leq +180.0°$.

Analogous to the bond-path curvature, see equation **(2)**, we may define dimensionless, *fractional* versions of the eigenvector-following path with length $\mathbb{H}$ where several forms are possible and not limited to the following:

$$\mathbb{H}_f = (\mathbb{H} - BPL)/BPL \qquad (4)$$

A similar expressions for $\mathbb{H}^*_f$ can be derived using the $\underline{e}_1$ eigenvector.

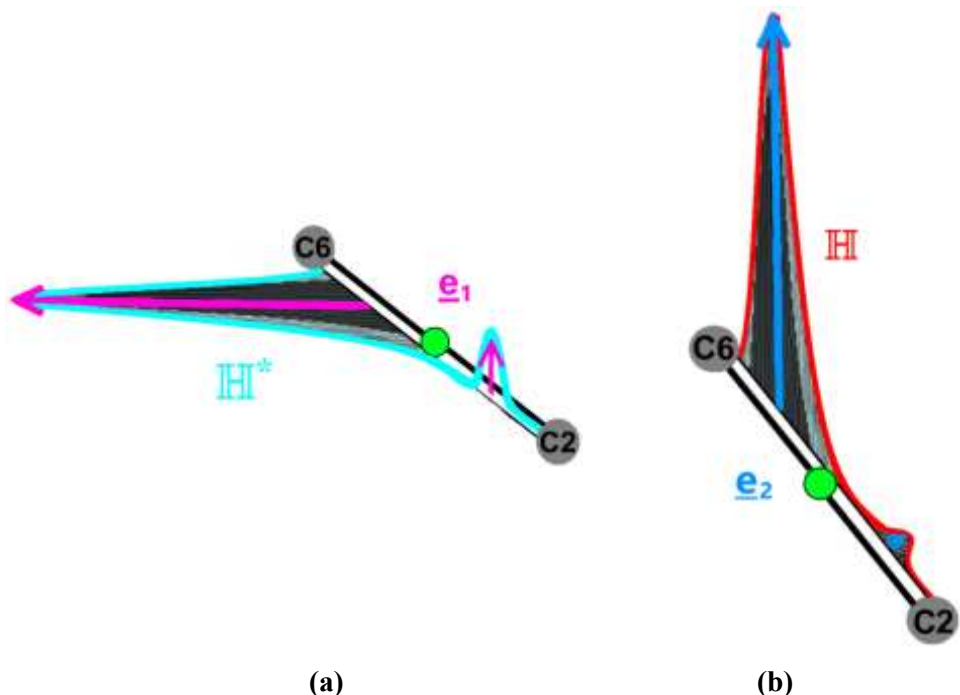

(a)           (b)

**Scheme 2.** The pale-blue line in sub-figure **(a)** represents the path, referred to as the eigenvector-following path with length $\mathbb{H}^*$, swept out by the tips of the scaled $\underline{e}_1$ eigenvectors, shown in magenta, and defined by equation **(2a)**. The red path in sub-figure **(b)** corresponds to the eigenvector-following path with length $\mathbb{H}$, constructed from the path swept out by the tips of the scaled $\underline{e}_2$ eigenvectors, shown in mid-blue and is defined by equation **(2b)**. The pale-blue and mid-blue arrows representing the $\underline{e}_1$ and $\underline{e}_2$ eigenvectors are scaled by the ellipticity ε respectively, where the vertical scales are exaggerated for visualization purposes. The green sphere indicates the position of a given *BCP*. Details of how to implement the calculation of the eigenvector-following paths with lengths $\mathbb{H}^*$ and $\mathbb{H}$ are provided in the **Supplementary Materials S8**.

The form of $\mathbb{H}_f$ defined by equation **(4a)** is the closest to the spirit of the bond-path curvature, equation **(2)**.

A bond within QTAIM is defined as being the bond-path traversed along the $\underline{e}_3$ eigenvector of the $\lambda_3$ eigenvalue from the bond-path, but, as a consequence of equation **(3)**, this definition should be expanded. This next-generation QTAIM definition of a bond should consider the bond-path to comprise the two paths swept out by the $\underline{e}_1$ and $\underline{e}_2$ eigenvectors that form the eigenvector-following path with length $\mathbb{H}^*$ and $\mathbb{H}$, respectively. Therefore, in this investigation we will consider the comparison with the stress tensor using the bond-path framework set $\mathbb{B}_{\sigma H} = \{p_{\sigma H}, q_{\sigma H}, r\}$ lengths $\mathbb{H}_{\sigma H}$ and $\mathbb{H}_{\sigma H}^*$ and also $\mathbb{B}_\sigma = \{p_\sigma, q_\sigma, r\}$ with corresponding lengths $\mathbb{H}_\sigma$ and $\mathbb{H}_\sigma^*$ using the definitions of the ellipticities $\varepsilon_{\sigma H}$ and $\varepsilon_\sigma$ defined by equation **1(a-b)** and equation **3(a-b)** respectively. The purpose of this comparison is two-fold, firstly to determine more definitively which of the $\underline{e}_{1\sigma}$ and $\underline{e}_{2\sigma}$ stress tensor eigenvectors correspond to the most/least preferred directions and secondly to determine the most useful stress tensor version of the bond-path framework set i.e. $\mathbb{B}_{\sigma H} = \{p_{\sigma H}, q_{\sigma H}, r\}$ or $\mathbb{B}_\sigma = \{p_\sigma, q_\sigma, r\}$ for use within the QTAIM partitioning.

## 3. Computational Details

The first step of the computational protocol is to perform a constrained scan of the potential energy surface of the C1-C2 *BCP*, see **Figure 1(a)** and **Figure 1(b)**. The scan was performed with a constrained (Z-matrix) geometry optimization performed at all steps with all coordinates free to vary except for the torsion coordinate θ. For the ethene molecule, the C2 end of the torsion C1-C2 *BCP* was held fixed, and the C1 end was rotated. The torsion coordinate θ was defined by the dihedral angle H3-C2-C1-H6 for the ethene in the range -180.0° ≤ θ ≤ +180.0° with 1.0° intervals and corresponding to clockwise(CW) and counter-clockwise(CCW) directions of torsion θ respectively. With tight convergence criteria at B3LYP/cc-pVQZ with Gaussian 09B01[23] was used. Subsequent single point energies for each step in the potential energy surface were evaluated using the same theory level, convergence criteria and integration grids. QTAIM and stress tensor analysis was performed with the AIMAll[24] suite on each wave function obtained in the previous step. The calculated paths comprising the $\mathbb{B}$, $\mathbb{B}_\sigma$ and $\mathbb{B}_{\sigma H}$ were visualized using the Python 3 visualization toolkit Mayavi[25].

## 4. Results and discussions

*4.1 A QTAIM and stress tensor BCP analysis of ethene*

In this section we present the *BCP* measures: the scalar ellipticity ε and the stress tensor ellipticities $\varepsilon_{\sigma H}$ and $\varepsilon_\sigma$ in addition to the vector stress tensor trajectory $\mathbb{T}_\sigma(s)$. The purpose of this is to determine the physical basis of the most preferred or 'easy' direction i.e. $\mathbf{e}_{1\sigma}$ or $\mathbf{e}_{2\sigma}$.

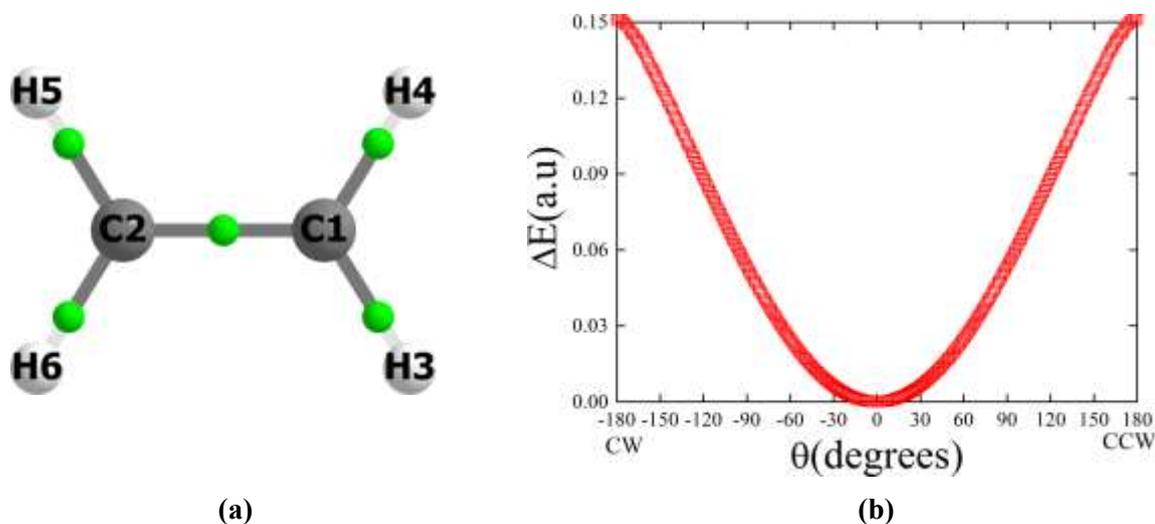

(a) (b)

**Figure 1**. The molecular graph of the ethene with the atom labelling scheme, with the green spheres indicate the *BCP*s, is presented in sub-figure **(a)**. The variation of the relative energy ΔE (in a.u.) of the ethene with the torsion θ, -180.0° ≤ θ ≤ +180.0° is presented in sub-figure **(b)**.

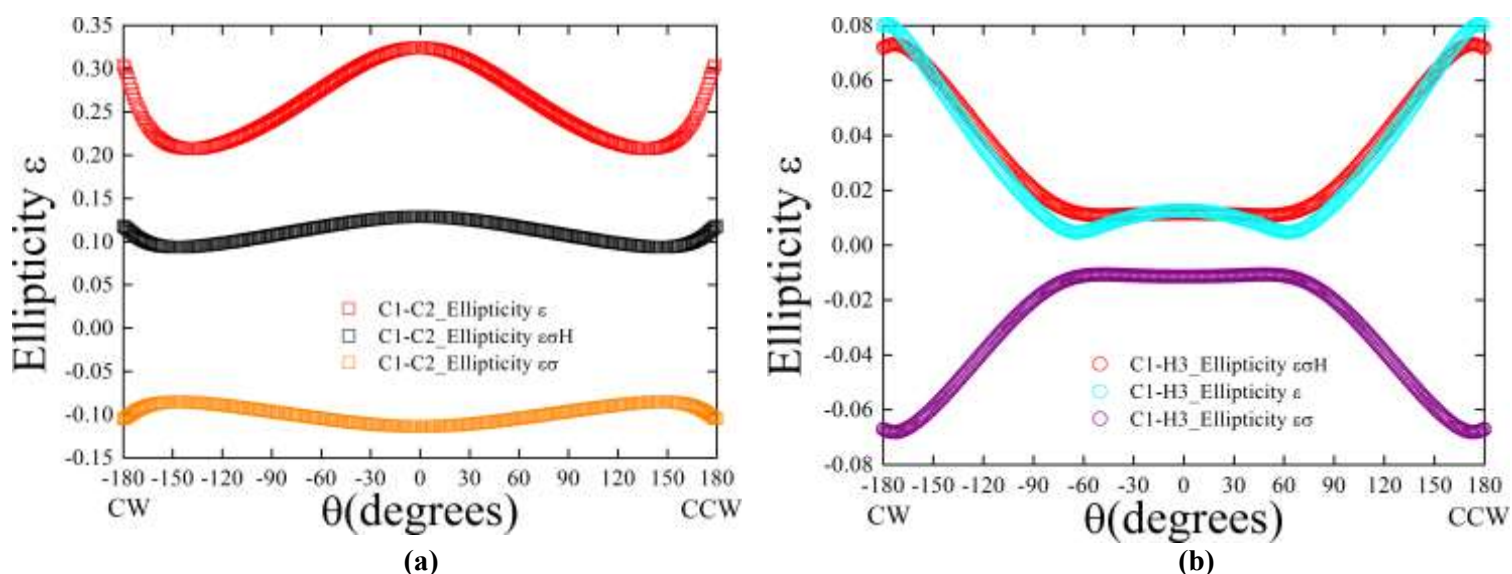

**Figure 2**. The variation of the three versions ellipticity $\varepsilon = |\lambda_1|/|\lambda_2| - 1$, $\varepsilon_{\sigma H} = |\lambda_{1\sigma}|/|\lambda_{2\sigma}| - 1$ and $\varepsilon_\sigma = |\lambda_{2\sigma}|/|\lambda_{1\sigma}| - 1$, for the C1-C2 *BCP* with the torsion θ are presented sub-figures **(a)**, see **Figure 1(a)** for the atom labelling scheme. The corresponding values for the C1-H3 *BCP* are presented in sub-figure **(b)**, the results for the C2-H6 *BCP* are provided in the **Supplementary Materials S1** results for stress tensor eigenvalue $\lambda_{3\sigma}$ are provided in the **Supplementary Materials S2**.

It can be seen that for values of the torsion θ ≈ ±150.0° that there is a change in the variation of the ellipticity ε of the C1-C2 *BCP* and C1-H3 *BCP* with torsion θ that is not indicated from examination of the variation of the relative energy ΔE, see **Figure 2** and **Figure 1** respectively. Therefore, due to the nature of the high degree of distortion of the C1-C2 *BCP* we will only consider the stress tensor trajectories $\mathbb{T}_\sigma(s)$ in the range -150.0° ≤ θ ≤ +150.0°, see **Figure 3**. We see that the forms of the variation of the ellipticity ε and the of the C1-H3 *BCP* is rather similar to that of the stress tensor ellipticity $\varepsilon_{\sigma H}$ are very similar that could mislead the reader into thinking that the *scalar* QTAIM and stress tensor behaviors are also similar in the general idea of using QTAIM as an approximation of the stress tensor, see **Figure 2(b)**. One would expect for a *symmetrical* bond-path that the corresponding QTAIM and stress tensor properties would be similar. Examination of the QTAIM and both versions of the stress tensor ellipticity $\varepsilon_{\sigma H}$ and $\varepsilon_\sigma$ for the symmetrical bond-path of the C1-C2 *BCP* however, shows that this not the case, see **Figure 2(a)**. Therefore we see that the scalar measure of the *BCP* ellipticity ε and ($\varepsilon_{\sigma H}$, $\varepsilon_\sigma$) are insufficient measures to compare the QTAIM and stress tensor schemes.

Therefore, we will consider a method to determine the 'easy' i.e. most preferred direction for the stress tensor using vector-based *BCP* measure the stress tensor trajectory $\mathbb{T}_\sigma(s)$. The applied torsion θ to the torsional C1-C2 *BCP* moves the ethene molecule away from the relative energy ΔE minimum, for both the CW and CCW torsion θ, see **Figure 1(b).** This is demonstrated by analysis of the stress tensor trajectory $\mathbb{T}_\sigma(s)$ of the torsional C1-C2 *BCP*, see **Figure 3**.

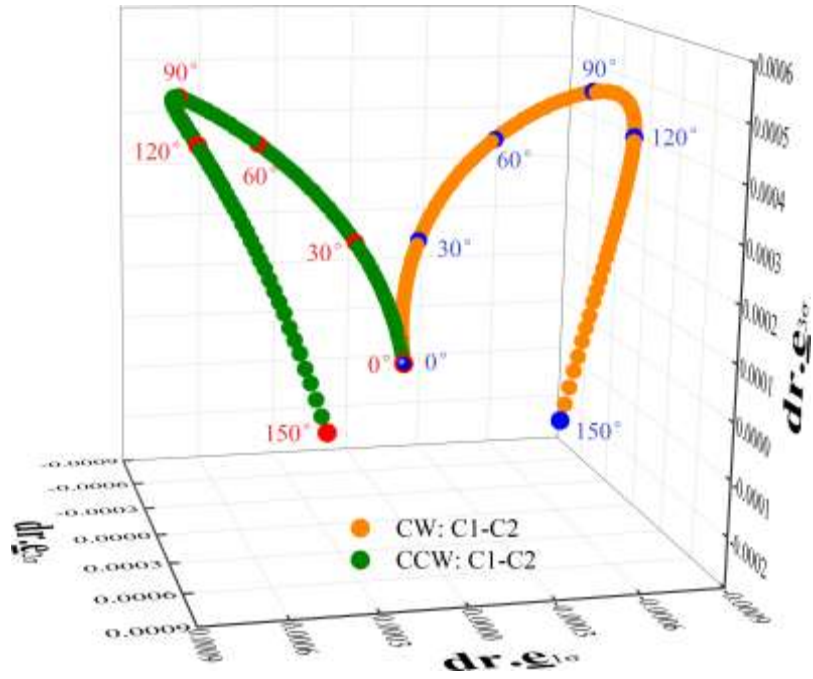

**Figure 3.** The stress tensor trajectories $\mathbb{T}_\sigma(s)$ of the torsional C1-C2 *BCP* in the eigenvector projection space $\mathbb{U}_\sigma(s)$ for the clockwise (CW) direction, $-150.0° \leq \theta \leq 0.0°$ and counterclockwise (CCW) direction, $0.0° \leq \theta \leq +150.0°$, corresponding trajectories $\mathbb{T}_\sigma(s)$ for the C1-H3 *BCP* and C2-H6 *BCP* are provided in **Table S3** and **Figure S2** of the **Supplementary Materials S3**.

The stress tensor trajectory $\mathbb{T}_\sigma(s)$ has a greater value of the maximum projection in the least preferred direction compared to the most direction, i.e. since $(\mathbf{e}_{2\sigma} \cdot \mathbf{dr})_{max} > (\mathbf{e}_{1\sigma} \cdot \mathbf{dr})_{max}$, see **Table 1**. This demonstrates that the stress tensor trajectory $\mathbb{T}_\sigma(s)$ is consistent with expectations from the potential energy surface in that the $\mathbf{e}_{1\sigma}$ eigenvector indicates the most preferred direction of electron motion. In other words, the $\mathbf{e}_{1\sigma}$ eigenvector indicates the most compressible direction as was discussed in the introduction.

**Table 1.** The maximum stress tensor trajectory $\mathbb{T}_\sigma$ projections $\{(\underline{\mathbf{e}}_{1\sigma} \cdot \mathbf{dr})_{max}, (\underline{\mathbf{e}}_{2\sigma} \cdot \mathbf{dr})_{max}, (\underline{\mathbf{e}}_{3\sigma} \cdot \mathbf{dr})_{max}\}$ for the CW direction of torsion $\theta$, where $-150.0° \leq \theta \leq 0.0°$. The corresponding results for the CCW torsion $\theta$ are identical are provided in the **Supplementary Materials S3**.

| *BCP* | $\{(\mathbf{e}_{1\sigma} \cdot \mathbf{dr})_{max}, (\mathbf{e}_{2\sigma} \cdot \mathbf{dr})_{max}, (\mathbf{e}_{3\sigma} \cdot \mathbf{dr})_{max}\}$ |
|---|---|
| **C1-C2** | **{8.250E-04, 1.000E-03, 4.840E-04}** |
| C1-H3 | {8.830E-03, 4.360E-04, 3.910E-03} |
| C2-H6 | {1.170E-02, 3.310E-02, 1.550E-02} |
| C1-H4 | {2.020E-03, 9.190E-03, 8.280E-03} |
| C2-H5 | {5.900E-03, 5.730E-03, 1.210E-02} |

*4.2 A QTAIM and stress tensor bond-path framework set analysis of ethene*

In this section we will determine which of the stress tensor ellipticities $\varepsilon_\sigma$ or $\varepsilon_{\sigma H}$ is the most useful for the purpose of using QTAIM to approximate the stress tensor and vice versa. This is will tested by creating versions of ($p_\sigma$-,$q_\sigma$-) or ($p_{\sigma H}$-,$q_{\sigma H}$-) paths to see which most closely resembles the QTAIM ($p$-,$q$-) paths.

The eigenvector following lengths ($\mathbb{H}^*$,$\mathbb{H}$), ($\mathbb{H}^*_{\sigma H}$,$\mathbb{H}_{\sigma H}$) and ($\mathbb{H}^*_\sigma$,$\mathbb{H}_\sigma$), associated with the ($p$-,$q$-), ($p_{\sigma H}$-,$q_{\sigma H}$-) and ($p_\sigma$-,$q_\sigma$-) paths are longer than the bond-path ($r$) for both the C1-C2 *BCP* bond-path, see **Figure 2** and **Figure 4** respectively. The eigenvector following lengths ($\mathbb{H}^*_{\sigma H}$,$\mathbb{H}_{\sigma H}$) and ($\mathbb{H}^*_\sigma$,$\mathbb{H}_\sigma$), associated with the ($p_\sigma$-,$q_\sigma$-) or ($p_{\sigma H}$-,$q_{\sigma H}$-) paths of the C-H *BCP*s however, can be shorter than the corresponding bond-path ($r$), see the middle and right panels of **Figure 4(b-c)** and theory section 2.2 for explanation. This seemingly anomalous effect is clearer for the fractional versions, see the middle and right panels of **Figure 5(b-c)**.

The variation with torsion $\theta$ of the stress tensor eigenvector following lengths ($\mathbb{H}^*_\sigma$,$\mathbb{H}_\sigma$) and ($\mathbb{H}^*_{\sigma H}$,$\mathbb{H}_{\sigma H}$) more closely follow the bond-path $r$ than do the corresponding QTAIM variations ($\mathbb{H}^*$,$\mathbb{H}$), compare the middle and right panels of **Figure 4(a)** with left panel of **Figure 4(a)**. The stress tensor ($\mathbb{H}^*_\sigma$,$\mathbb{H}_\sigma$) version being the most similar to the bond-path and for the C1-C2 *BCP* the $\mathbb{H}^*_\sigma$ and $\mathbb{H}_\sigma$ values are somewhat indistinguishable from each other as are corresponding values for the $\mathbb{H}^*_{\sigma H}$ and $\mathbb{H}_{\sigma H}$. Therefore, the variations with the torsion $\theta$ for the C1-C2 *BCP* of neither of the scalar stress tensor lengths ($\mathbb{H}^*_\sigma$,$\mathbb{H}_\sigma$) and ($\mathbb{H}^*_{\sigma H}$,$\mathbb{H}_{\sigma H}$) are a good approximation for the behavior of the QTAIM lengths ($\mathbb{H}^*$,$\mathbb{H}$).

The value of $\mathbb{H}^* < \mathbb{H}$ for the C1-C2 *BCP* away from the relaxed geometry see the left panel of **Figure 4(a)** and $\mathbb{H}^* < \mathbb{H}$ for all values of the torsion $\theta$ for the C1-H3 *BCP* and C2-H6 *BCP*, see the left panels of **Figure 4(b)** and **Figure 4(c)** respectively. This indicates that for QTAIM the preferred $\underline{e}_2$ direction is associated with the longer path $\mathbb{H}$, from the form of $q_i = r_i + \varepsilon_i\underline{e}_{2,i}$ and equation **(3b)**.

The asymmetry of the positions of the C1-H3 *BCP* and C2-H6 *BCP* on the corresponding bond-paths arises due to the mismatch in the positions of the $\nabla\cdot\rho(\mathbf{r_b}) = 0$ and $-\nabla\cdot\sigma(\mathbf{r_b}) = 0$ associated with the QTAIM and stress tensor respectively. The consequences of this asymmetry are apparent from the presence of the asymmetrical variations of the $\mathbb{H}_\sigma$ and $\mathbb{H}_{\sigma H}$ of the C1-H3 *BCP* and C2-H6 *BCP*, see the middle and right panels of **Figure 4(b)** and **Figure 4(c)** respectively. Conversely, the corresponding variations of the $\mathbb{H}^*_\sigma$ and $\mathbb{H}^*_{\sigma H}$ with the torsion $\theta$ for the C1-H3 *BCP* and C2-H6 *BCP* are symmetrical following the form of the bond-path ($r$) and are generally longer than those of $\mathbb{H}_\sigma$ and $\mathbb{H}_{\sigma H}$. The significance of this is that for the stress tensor the most preferred directions are those associated with $\mathbb{H}^*_\sigma$ and $\mathbb{H}^*_{\sigma H}$, because they are constructed from the most preferred $\underline{e}_{1\sigma}$ the direction that is associated with the highest degree of compressibility.

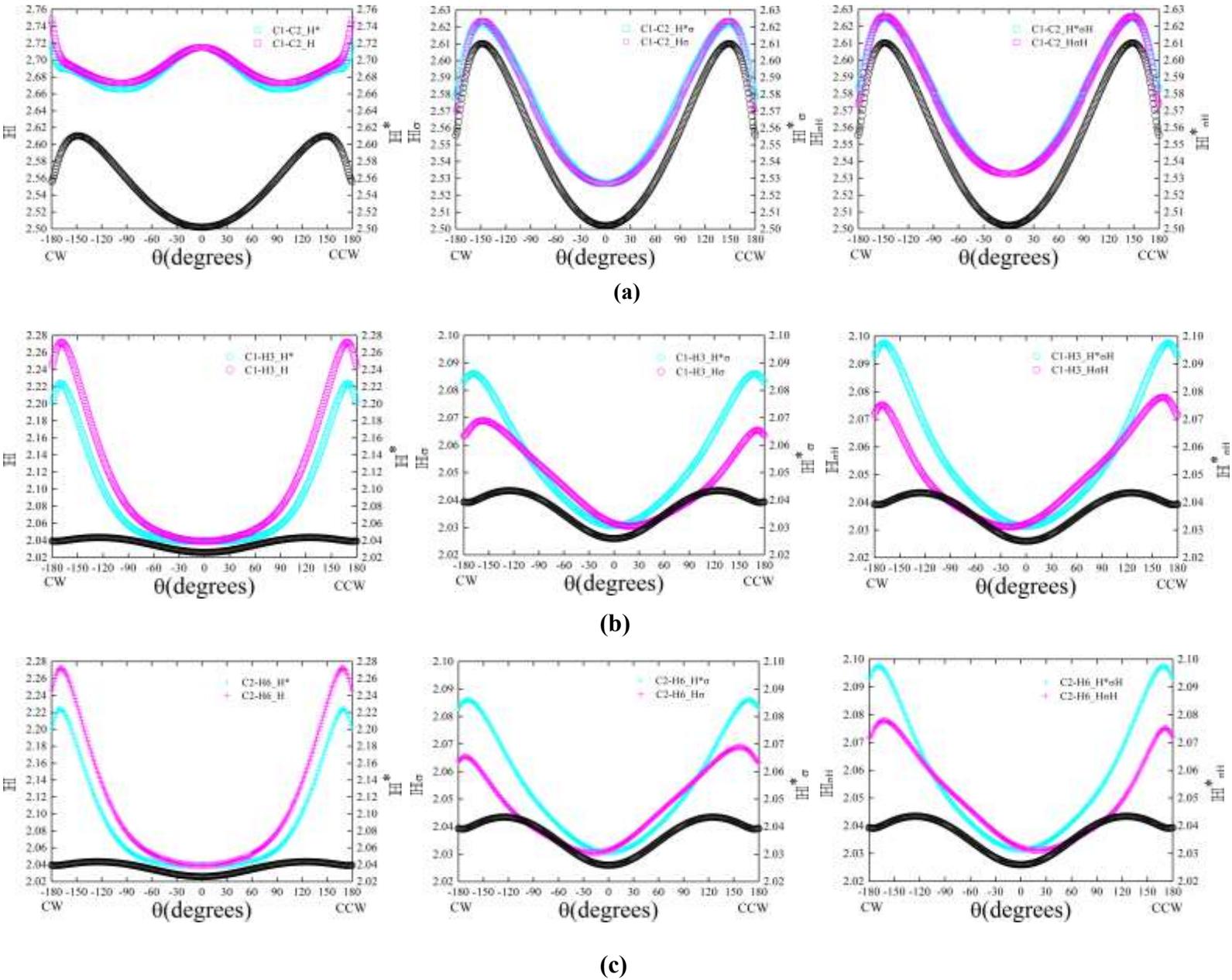

**Figure 4**. The variation of the eigenvector-following path length of the C1-C2 *BCP* with torsion θ: $\mathbb{H}$, $\mathbb{H}_\sigma$ and $\mathbb{H}_{\sigma H}$, are denoted by the pale-blue plot lines in the left, middle and right panels respectively. The corresponding values for $\mathbb{H}^*$, $\mathbb{H}^*_\sigma$ and $\mathbb{H}^*_{\sigma H}$ and the bond-path lengths (BPL) are denoted by the magenta and black plot lines respectively, also see **Figure 1(a)** for the atom labelling scheme. The corresponding plots for the C1-H3 *BCP* and C2-H6 *BCP* are presented in sub-figure **(b)** and sub-figure **(c)** respectively.

Examination of the fractional versions ($\mathbb{H}^*_f, \mathbb{H}_f$) of the eigenvector-following path for the stress tensor adapted from equation (**4**) as ($\mathbb{H}^*_{f\sigma}, \mathbb{H}_{f\sigma}$) and ($\mathbb{H}^*_{f\sigma H}, \mathbb{H}_{f\sigma H}$). These fractional versions demonstrate the preference of the $\underline{e}_{1\sigma}$ direction for the C1-C2 *BCP* since for most values of the torsion θ, $\mathbb{H}^*_\sigma > \mathbb{H}_\sigma$ and also $\mathbb{H}^*_{\sigma H} > \mathbb{H}_{\sigma H}$, see left panel of **Figure 5(a)**. We again see that the $\underline{e}_{1\sigma}$ direction is indicated as preferred for the C1-H3 *BCP* and C2-H6 *BCP* and that the corresponding variation of the $\mathbb{H}^*_\sigma$ and $\mathbb{H}^*_{\sigma H}$ with the torsion θ is symmetrical, see the middle and right panels of **Figure 5(b)** and **Figure 5(c)** respectively.

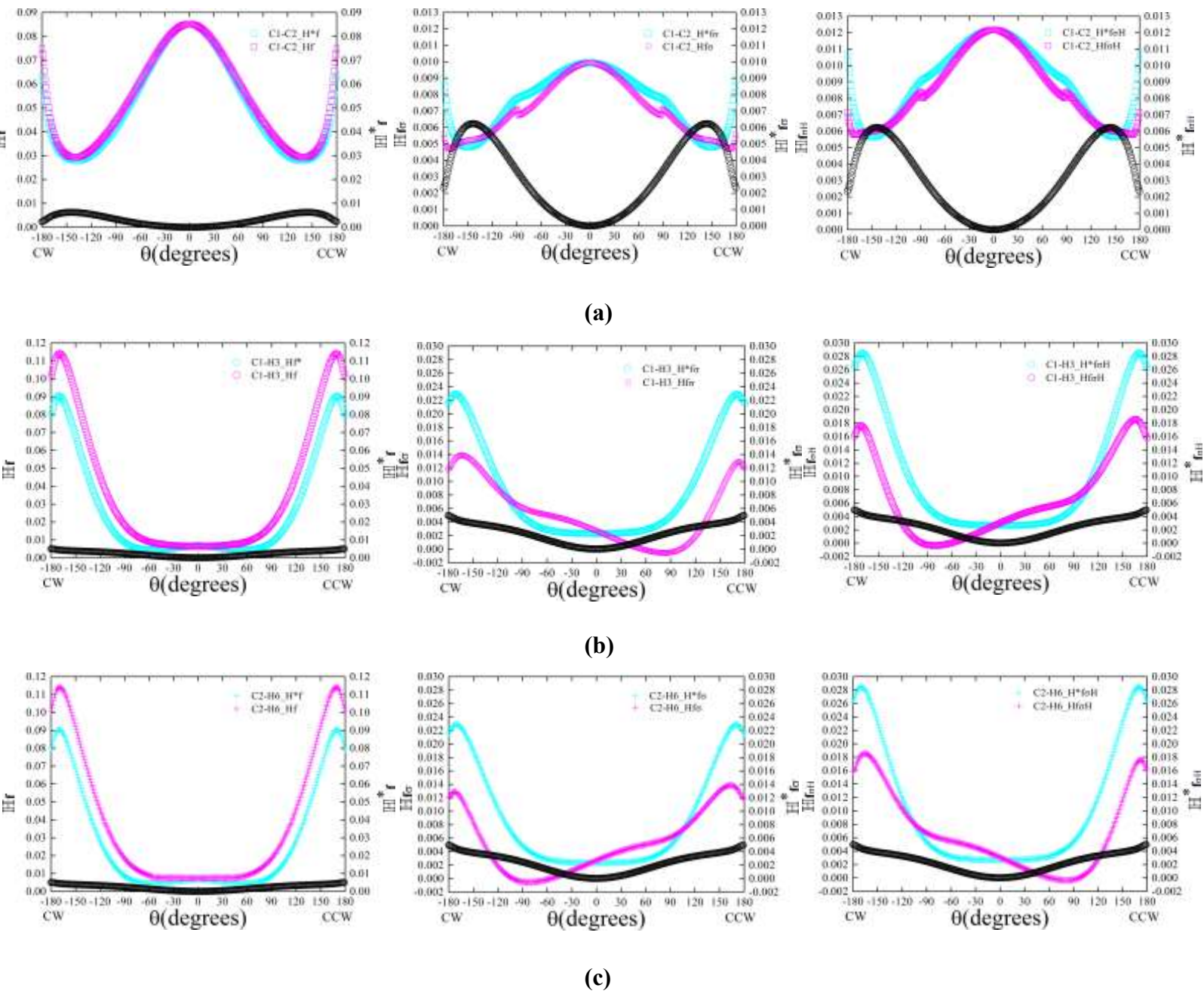

**Figure 5.** The variation of the fractional eigenvector-following path lengths $\mathbb{H}_f$, $\mathbb{H}_{\sigma f}$ and $\mathbb{H}_{\sigma Hf}$ of the C1-C2 *BCP*, C1-H3 *BCP* and C2-H6 *BCP* with torsion θ are shown in sub-figures **(a-c)** respectively, the bond-path curvatures are denoted by the black plot lines, see also the caption of **Figure 3.** The corresponding results for $\mathbb{H}_{fmin}$, $\mathbb{H}^*_{fmin}$, $\mathbb{H}_{fmin\sigma}$, $\mathbb{H}^*_{fmin\sigma}$, $\mathbb{H}_{fmin\sigma H}$, $\mathbb{H}^*_{fmin\sigma H}$ are provided in the **Supplementary Materials S4**.

Comparison of the QTAIM bond-path framework set $\mathbb{B} = \{p,q,r\}$ with the two stress tensor variants $\mathbb{B}_\sigma = \{p_\sigma,q_\sigma,r\}$ and $\mathbb{B}_{\sigma H} = \{p_{\sigma H},q_{\sigma H},r\}$ demonstrates that $\mathbb{B}_\sigma = \{p_\sigma,q_\sigma,r\}$ most closely resembles the QTAIM $\mathbb{B} = \{p,q,r\}$ particularly for the bond-path associated with torsional C1-C2 *BCP*, compare the left, middle and right panels of **Figure 6**. It can be seen that only considering the values of the stress tensor at the *BCP* to be misleading and provides an incomplete understanding of the behavior of the eigenvectors in the form of the QTAIM (*p-*,*q-*) and stress tensor (*p_σ-*,*q_σ-*) and (*p_σH-*,*q_σH-*) paths. Inspection of these paths in the vicinity of the C-H *BCP*s could indicate that the stress tensor (*p_σH-*,*q_σH-*) paths most closely resemble those of QTAIM

(*p*-,*q*-) paths, when instead consideration of the entire bond-path r shows that the (*p*$_\sigma$-,*q*$_\sigma$-) paths most closely resemble the QTAIM (*p*-,*q*-) paths. Results for individual (*p*-,*q*-) (*p*$_\sigma$-,*q*$_\sigma$-) and (*p*$_{\sigma H}$-,*q*$_{\sigma H}$-) paths are provided in the **Supplementary Materials S6**. We see that the resemblance of the stress tensor (*p*$_\sigma$-,*q*$_\sigma$-) paths with QTAIM (*p*-,*q*-) paths is maintained for the duration of the -150.0° ≤ θ ≤ +150.0°, this demonstrates the robustness of the approximation, see the left and middle panels of **Figure 6** and the **Supplementary Materials S5**.

Examination of the ellipticity $\varepsilon$, $\varepsilon_\sigma$ and $\varepsilon_{\sigma H}$ profiles for the C1-C2 *BCP* the ethene along the bond-path (*r*) shows a decrease in ± magnitude and a narrowing of the decrease in the $\varepsilon$, $\varepsilon_\sigma$ and $\varepsilon_{\sigma H}$ profiles with increase in the applied torsion θ, see **Figure 7(a)**. The variation of the ellipticity $\varepsilon$, $\varepsilon_\sigma$ and $\varepsilon_{\sigma H}$ profiles for the C1-H3 *BCP* and C2-H6 *BCP* maintains a similar width with increase in the applied torsion θ however, corresponding amplitude ± of the $\varepsilon$, $\varepsilon_\sigma$ and $\varepsilon_{\sigma H}$ profiles increases, see **Figure 7(b)**. We also notice that the peak values in the variation of the bond-path ellipticity $\varepsilon$, $\varepsilon_\sigma$ and $\varepsilon_{\sigma H}$ profiles for the C1-H3 *BCP* and C2-H6 *BCP* significantly away from the position of the *BCP* and closer to the C nuclei, see **Figure 7(b)**.

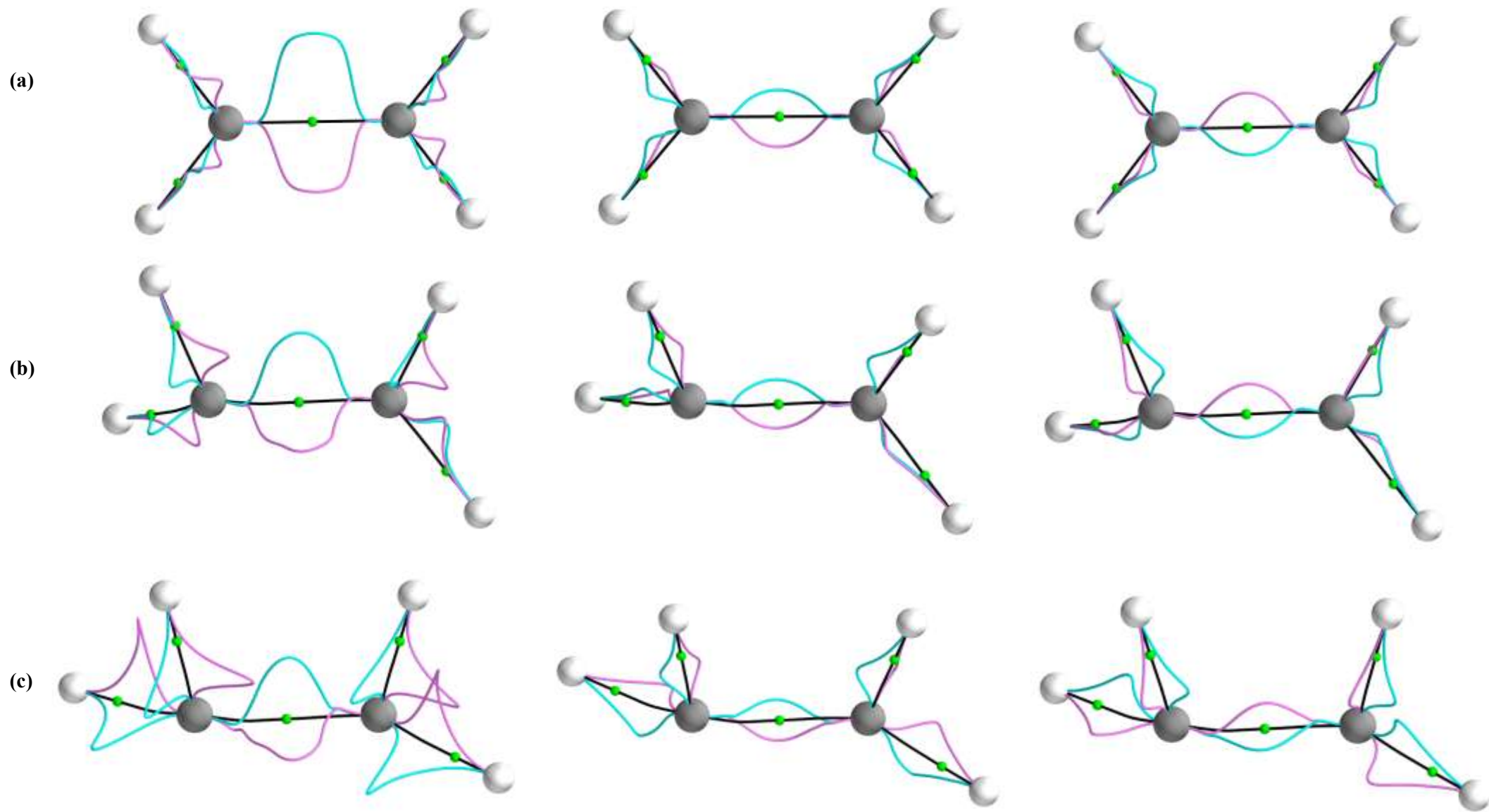

**Figure 6.** The bond-path framework sets $\mathbb{B} = \{p,q,r\}$, $\mathbb{B}_\sigma = \{p_\sigma,q_\sigma,r\}$ and $\mathbb{B}_{\sigma H} = \{p_{\sigma H},q_{\sigma H},r\}$ showing magnified (x5) versions of the ($p$-,$q$-), ($p_\sigma$-,$q_\sigma$-) and ($p_{\sigma H}$-,$q_{\sigma H}$-) paths are presented in the left, middle and right panels in sub-figures **(a)-(c)** respectively. The $p$-, $p_\sigma$- and $p_{\sigma H}$-paths (pale-blue) and $q$-, $q_\sigma$- and $q_{\sigma H}$-paths (magenta) and the $r$-path i.e. bond-path (black) corresponding ethene rotated in the clockwise (CW) direction for values of the torsion θ = 0.0°, 90.0° and 150.0° are presented in sub-figures **(a)-(c)** respectively. The corresponding results for the counterclockwise (CCW) direction are identical, are provided in the **Supplementary Materials S5**.

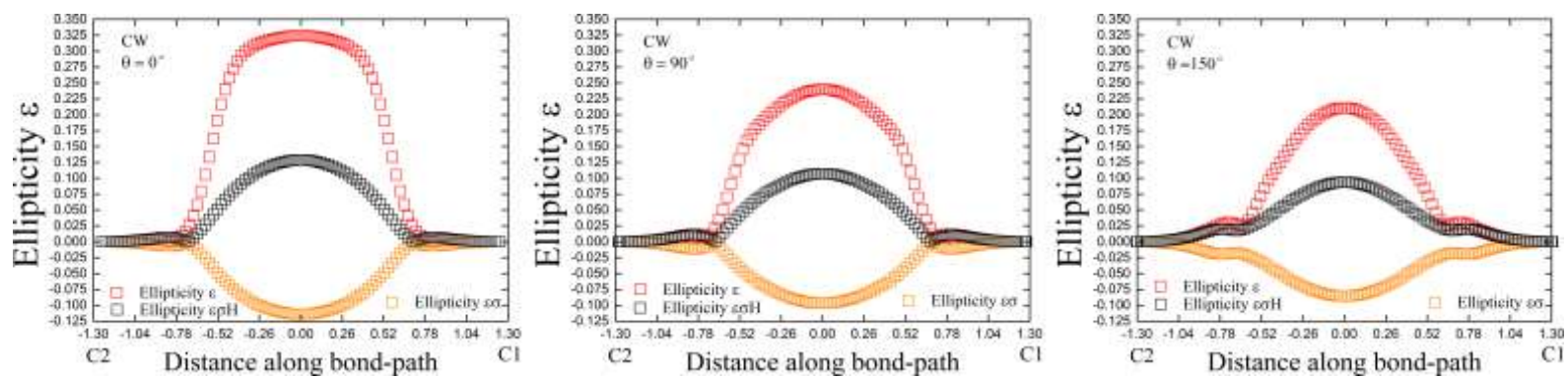

(a)

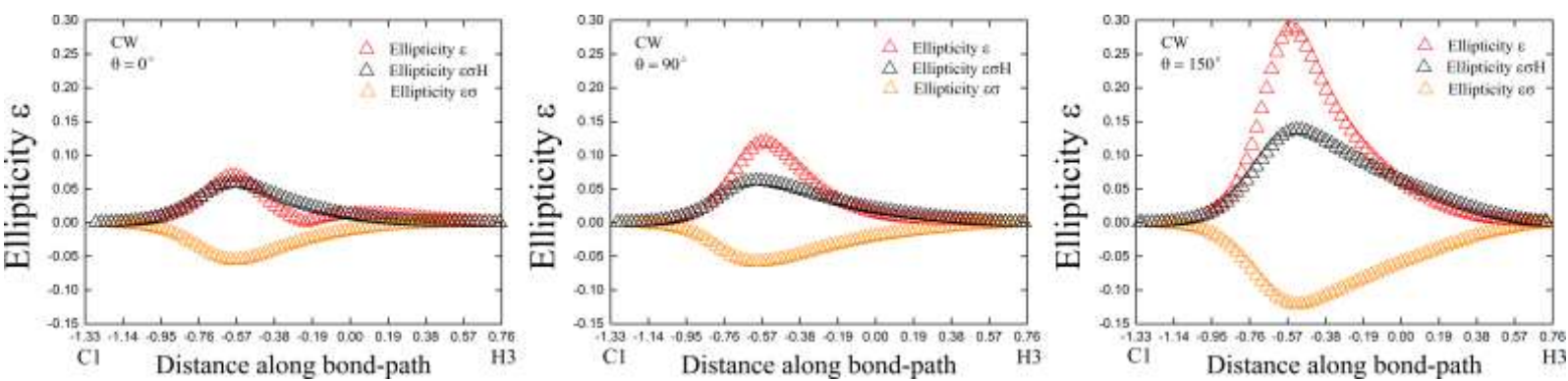

(b)

**Figure 7.** The variations of the three ellipticity ε, ε$_{\sigma H}$ and ε$_\sigma$ profiles for the ethene in clockwise (CW) directions along the bond-path (*r*) associated with the C1-C2 *BCP* and C1-H3 *BCP*, where θ = 0.0°, 90.0°, 150.0° are presented in sub-figures **(a)-(c)** respectively. The plots corresponding to θ = 30.0°, 60.0° and 120.0° are provided in **Supplementary Materials S7**.

**Conclusions**

We have demonstrated for the first time that the stress tensor can be used within the QTAIM partitioning scheme using an adapted version of the recently introduced 3-D *vector-based* interpretation of the chemical bond $\mathbb{B} = \{p,q,r\}$ of the form $\mathbb{B}_\sigma = \{p_\sigma,q_\sigma,r\}$ to follow changes in the directional properties of the stress tensor that is robust to large torsions. This three-stranded interpretation of the bond is more complete than minimal definition of bonding (**e**$_3$) provided by the bond-path (*r*) because it comprises all three of the {**e**$_1$, **e**$_2$, **e**$_3$} eigenvectors. Within QTAIM the most preferred 'easy' direction **e**$_2$ is determined on the basis of the ease of total electronic charge density $\rho(\mathbf{r_b})$ accumulation. Conversely, the least preferred directions were found to be **e**$_1$ for QTAIM and **e**$_{2\sigma}$ for the stress tensor. For the stress tensor we have found that the **e**$_{1\sigma}$ eigenvector corresponds to the most preferred 'easy' direction on the basis of ease of compressibility. This finding was demonstrated using stress tensor trajectory formalism $\mathbb{T}_\sigma(s)$ in partnership with the potential energy surface to prove that the **e**$_{2\sigma}$ eigenvector was the least preferred direction of electronic charge density $\rho(\mathbf{r_b})$ accumulation and therefore that the **e**$_{1\sigma}$ eigenvector was the most preferred direction. Additional evidence for the most and least preferred directions for the stress tensor being defined by the $\underline{e}_{1\sigma}$ and $\underline{e}_{2\sigma}$ eigenvector was provided by the fact that the values for $\mathbb{H}_\sigma^* > \mathbb{H}_\sigma$ and $\mathbb{H}_{\sigma H}^* > \mathbb{H}_{\sigma H}$ consistent with previous findings from QTAIM that the preferred path has the longer associated eigenvector following path length. A new indication from this work, applicable to the stress tensor, is that the least preferred eigenvector is indicated by the presence of asymmetrical variations of $\mathbb{H}_\sigma$ and $\mathbb{H}_{\sigma H}$ with the applied torsion θ.

Examination of the scalar ellipticity ε, $\varepsilon_{\sigma H}$ and $\varepsilon_\sigma$ demonstrated the insufficiency of any scalar measure for use in any electron proceeding method due to the lack of directional information provided by a scalar. Instead, we have presented the QTAIM bond-path framework set $\mathbb{B} = \{p,q,r\}$ and the stress tensor versions, $\mathbb{B}_\sigma = \{p_\sigma,q_\sigma,r\}$ and $\mathbb{B}_{\sigma H} = \{p_{\sigma H},q_{\sigma H},r\}$ where each of the three constituent paths are vector quantities that display the network of most preferred and least preferred directions of motion of the associated bond-paths and *BCP*s. We suggest that the construction of $\mathbb{B}_\sigma = \{p_\sigma,q_\sigma,r\}$ that uses $\varepsilon_\sigma \leq 0$ is more useful than the commonly used construction of the stress tensor ellipticity $\varepsilon_\sigma \geq 0$ because for $\mathbb{B}_\sigma = \{p_\sigma,q_\sigma,r\}$ the $p_\sigma$ and $q_\sigma$ paths more closely resemble the *p* and *q* from QTAIM than the $p_{\sigma H}$- and $q_{\sigma H}$-path. The reason for the counterintuitive result that $\varepsilon_\sigma \leq 0$ is more useful is because the 'easy' direction for the stress tensor is determined by the most compressible $\lambda_{1\sigma}$ i.e. associated with the longer axis of the ellipse, whereas for QTAIM the 'easy' direction, (longer axis of the ellipse) is associated with the $\lambda_2$ eigenvalue,.

In addition, we have demonstrated the importance of considering a 3-D vector-based measure of bonding that can follow the entire bond-path as opposed to only at the *BCP* when QTAIM to obtain the stress tensor properties. This was demonstrated by the fact that the $p_\sigma$-,$q_\sigma$-paths and $p_{\sigma H}$-,$q_{\sigma H}$-paths twist about the *BCP* for the asymmetrical C-H *BCP* bond-path *r* and the ellipticity profiles ε, $\varepsilon_\sigma$ and $\varepsilon_{\sigma H}$ profiles for the C1-H3 and C2-H6 *BCP* display peak values well away from the location of the associated *BCP*s.

Therefore, despite the fact that the stress tensor *partitioning* is not generally available as a standard output option in computational software or computable from an experimentally measured electron density we have shown that we can use the stress tensor results obtained within the QTAIM partitioning.

Future work includes using the bond-path framework set $\mathbb{B}_\sigma = \{p_\sigma,q_\sigma,r\}$ within the QTAIM partitioning for more complex reactions, starting with photo-isomerization reactions, currently in progress, followed by ring-opening reactions, $S_N2$ reactions and ESIPT reactions. Related work could also include the creation of an Ehrenfest Force **F(r)** partitioning bond-path framework set $\mathbb{B}_F = \{p_F,q_F,r_F\}$ with a complete Ehrenfest Force **F(r)** molecular graph including the bond-paths $r_F$.

**Acknowledgements**

The National Natural Science Foundation of China is gratefully acknowledged, project approval number: 21673071. The One Hundred Talents Foundation of Hunan Province and the aid program for the Science and Technology Innovative Research Team in Higher Educational Institutions of Hunan Province are also gratefully acknowledged for the support of S.J. and S.R.K.

# SUPPLEMENTARY MATERIALS

# Stress Tensor Eigenvector Following with Next-Generation Quantum Theory of Atoms in Molecules

Jia Hui Li, Wei Jie Huang, Tianlv Xu, Steven R. Kirk[*] and Samantha Jenkins[*]

*Key Laboratory of Chemical Biology and Traditional Chinese Medicine Research and Key Laboratory of Resource Fine-Processing and Advanced Materials of Hunan Province of MOE, College of Chemistry and Chemical Engineering, Hunan Normal University, Changsha, Hunan 410081, China*

email: steven.kirk@cantab.net
email: samanthajsuman@gmail.com

**1. Supplementary Materials S1.** The variation of the three versions ellipticity ε for the the C2-H6 *BCP*.

**2. Supplementary Materials S2.** The variation of the stress tensor eigenvalue $\lambda_{3\sigma}$ of the C1-C2 *BCP*, C1-H3 *BCP* and C2-H6 *BCP* with the torsion θ.

**3. Supplementary Materials S3.** The maximum stress tensor $\mathbb{U}_\sigma$ space projections for the CCW direction and stress tensor trajectories $\mathbb{T}_\sigma(s)$.

**4. Supplementary Materials S4.** The variation of the fractional lengths $\mathbb{H}_{fmin}$, $\mathbb{H}^*_{fmin}$, $\mathbb{H}_{fmin\sigma}$, $\mathbb{H}^*_{fmin\sigma}$, $\mathbb{H}_{fmin\sigma H}$, $\mathbb{H}^*_{fmin\sigma H}$ with the torsion θ.

**5. Supplementary Materials S5.** Magnified (x5) versions of the (*p*,*q*), ($p_\sigma$, $q_\sigma$) and ($p_{\sigma H}$, $q_{\sigma H}$) paths along the bond-path (*r*) in the clockwise (CW) and the counterclockwise (CCW) directions.

**6. Supplementary Materials S6.** The (*p*,*q*), ($p_\sigma$,$q_\sigma$) and ($p_{\sigma H}$,$q_{\sigma H}$) paths along the bond-path (*r*) in the CW and CCW directions.

**7. Supplementary Materials S7.** The variations of the three ellipticity ε profiles in the CW direction along the bond-path (*r*) associated with the C1-C2 *BCP*, C1-H3 *BCP* and C2-H6 *BCPs*.

**8. Supplementary Materials S8.** Implementation details of the calculation of the eigenvector-following path lengths $\mathbb{H}$, $\mathbb{H}^*$ and $\mathbb{H}_\sigma^*$, $\mathbb{H}_{\sigma H}^*$.

# 1. Supplementary Materials S1.

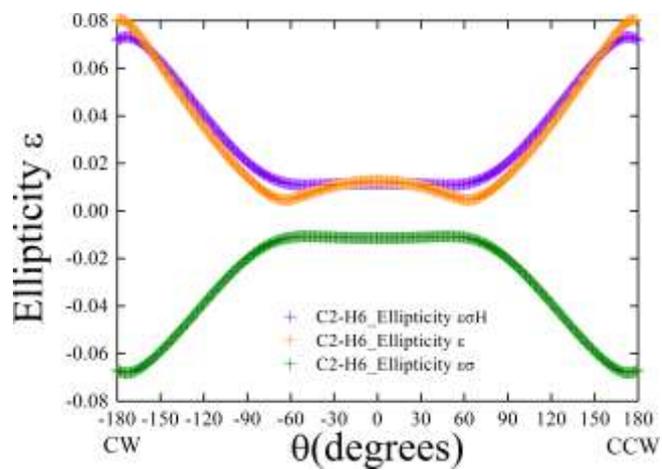

**Figure S1.** The variation of the three versions ellipticity $\varepsilon = |\lambda_1|/|\lambda_2| - 1$, $\varepsilon_\sigma = |\lambda_{2\sigma}|/|\lambda_{1\sigma}| - 1$, $\varepsilon_{\sigma H} = |\lambda_{1\sigma}|/|\lambda_{2\sigma}| - 1$ for the the C2-H6 *BCP*.

## 2. Supplementary Materials S2.

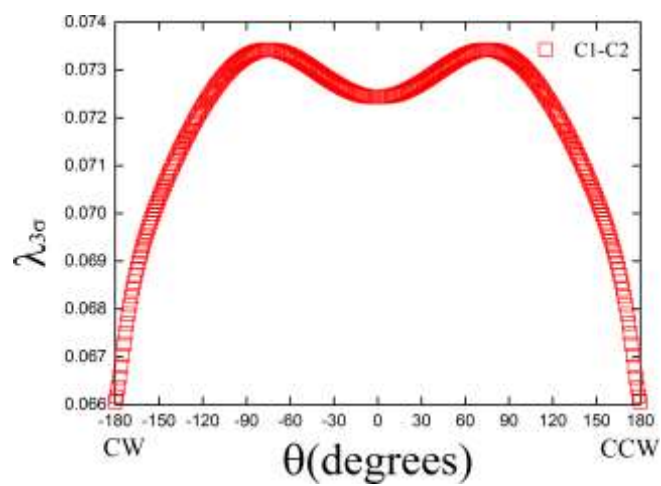

(a)

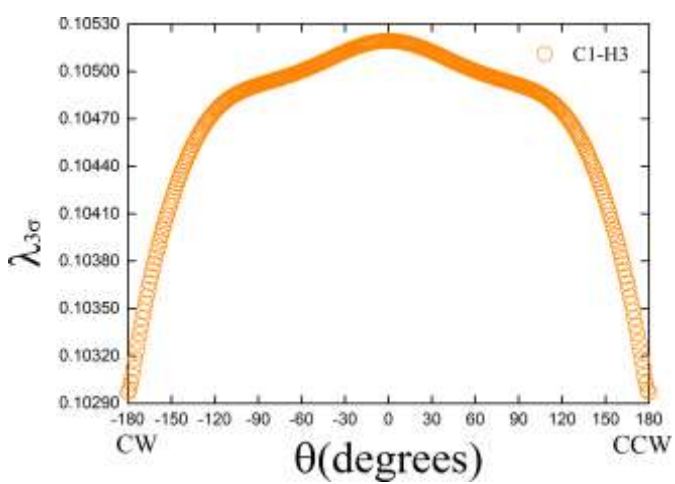

(b)

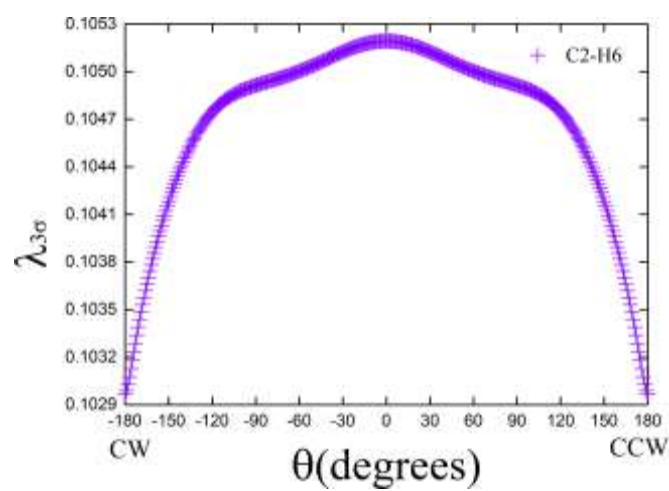

(c)

**Figure S2.**. The variation of the stress tensor eigenvalue $\lambda_{3\sigma}$ of the C1-C2 *BCP*, C1-H3 *BCP* and C2-H6 *BCP* with the torsion $\theta$ for the ethene are presented in sub-figures **(a-c)** respectively.

## 3. Supplementary Materials S3.

**Table S3.** The maximum stress tensor $\mathbb{U}_\sigma$ space projections $\{(\underline{e}_{1\sigma} \cdot dr)_{max}, (\underline{e}_{2\sigma} \cdot dr)_{max}, (\underline{e}_{3\sigma} \cdot dr)_{max}\}$ for the ethene at CCW directions. The maximum projections $\{(\underline{e}_{1\sigma} \cdot dr)_{max}, (\underline{e}_{2\sigma} \cdot dr)_{max}, (\underline{e}_{3\sigma} \cdot dr)_{max}\}$ for the torsion C1-C2 *BCP* are shown highlighted in a bold font.

|       | $\{(e_{1\sigma} \cdot dr)_{max}, (e_{2\sigma} \cdot dr)_{max}, (e_{3\sigma} \cdot dr)_{max}\}$ CCW |
|-------|---|
| BCP   |   |
| **C1-C2** | **{8.250E-04, 1.000E-03, 4.840E-04}** |
| C1-H3 | {8.830E-03, 4.360E-04, 3.910E-03} |
| C1-H4 | {2.020E-03, 9.190E-03, 8.280E-03} |
| C2-H5 | {5.900E-03, 5.730E-03, 1.210E-02} |
| C2-H6 | {1.170E-02, 3.310E-02, 1.550E-02} |

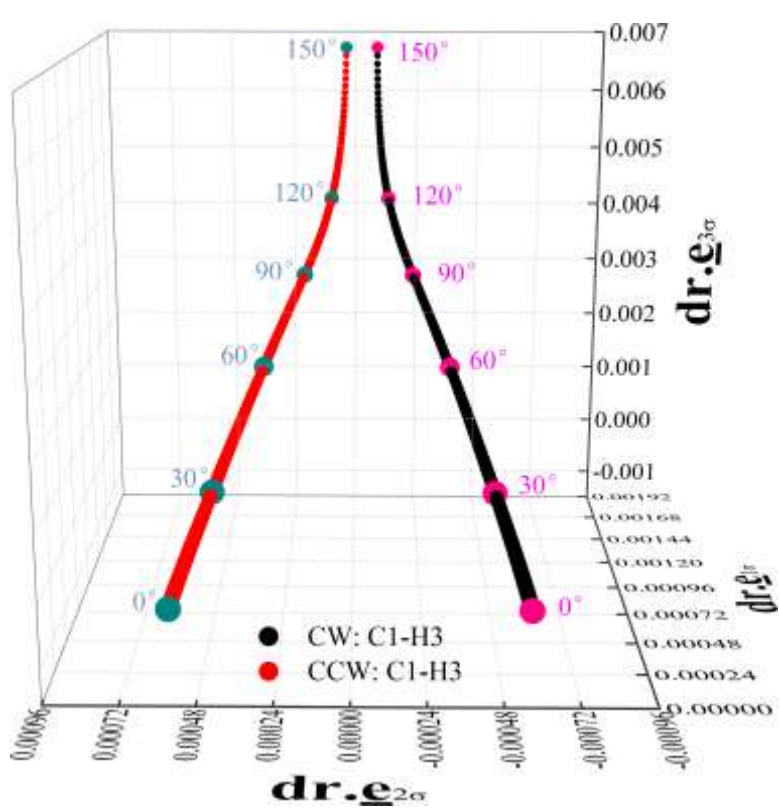

(a)

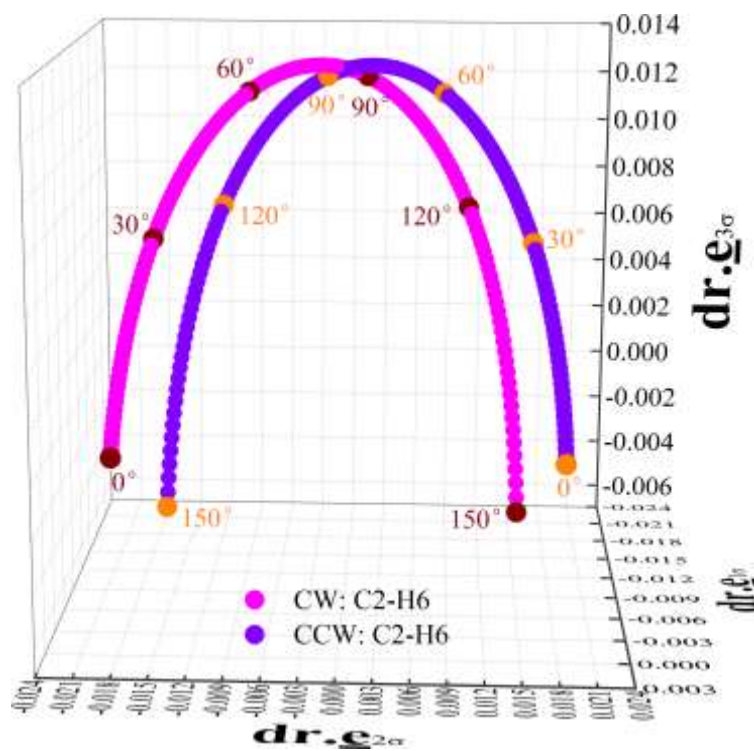

(b)

**Figure S3.** The stress tensor trajectories $\mathbb{T}_\sigma(s)$ for the clockwise (CW) direction, $-150.0° \leq \theta \leq 0.0°$ and counterclockwise (CCW) direction, $0.0° \leq \theta \leq +150.0°$, corresponding trajectories $\mathbb{T}_\sigma(s)$ for the C1-H3 *BCP* and C2-H6 *BCP* are shown in sub-figures **(a)** and **(b)** respectively.

## 4. Supplementary Materials S4.

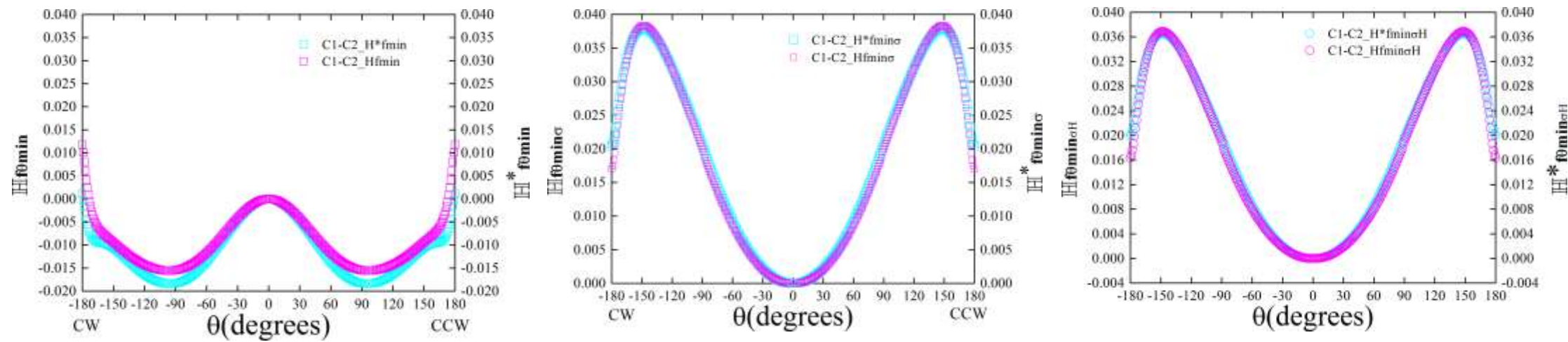

(a)

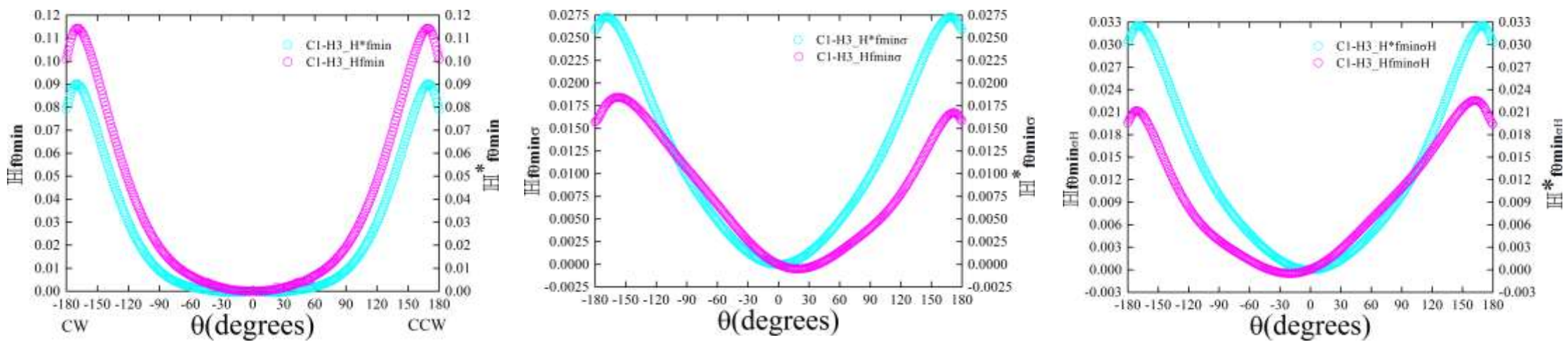

(b)

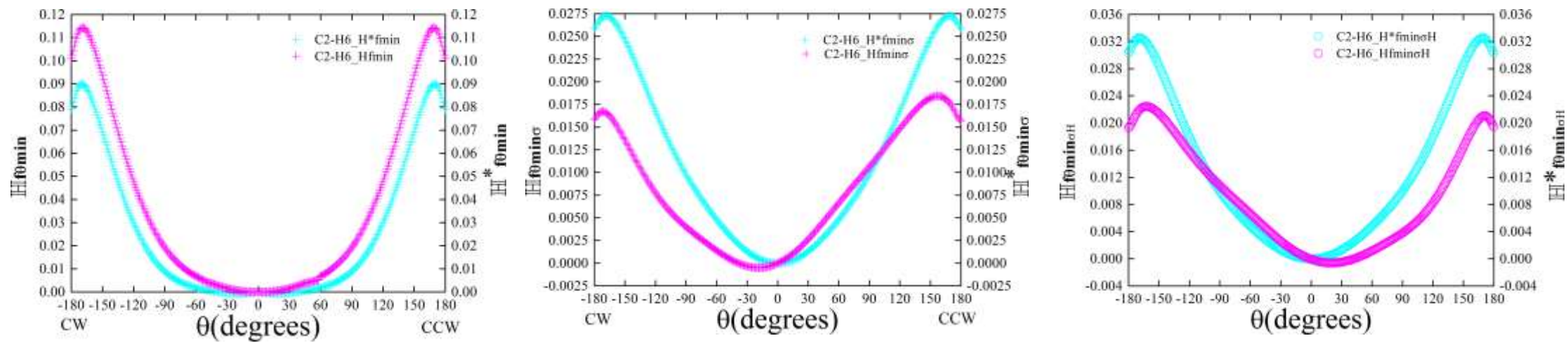

**(c)**

**Figure S4.** The variation of the eigenvector-following path length $\mathbb{H}_{fmin}$, $\mathbb{H}^*_{fmin}$, $\mathbb{H}_{fmin\sigma}$, $\mathbb{H}^*_{fmin\sigma}$, $\mathbb{H}_{fmin\sigma H}$, $\mathbb{H}^*_{fmin\sigma H}$ of the C1-H2 *BCP*, C1-H3 *BCP* , C2-H6 *BCP* with torsion θ are denoted by the magenta, pale-blue are shown in the left, middle and right panels of sub-figures **(a-c)** respectively.

## 5. Supplementary Materials S5.

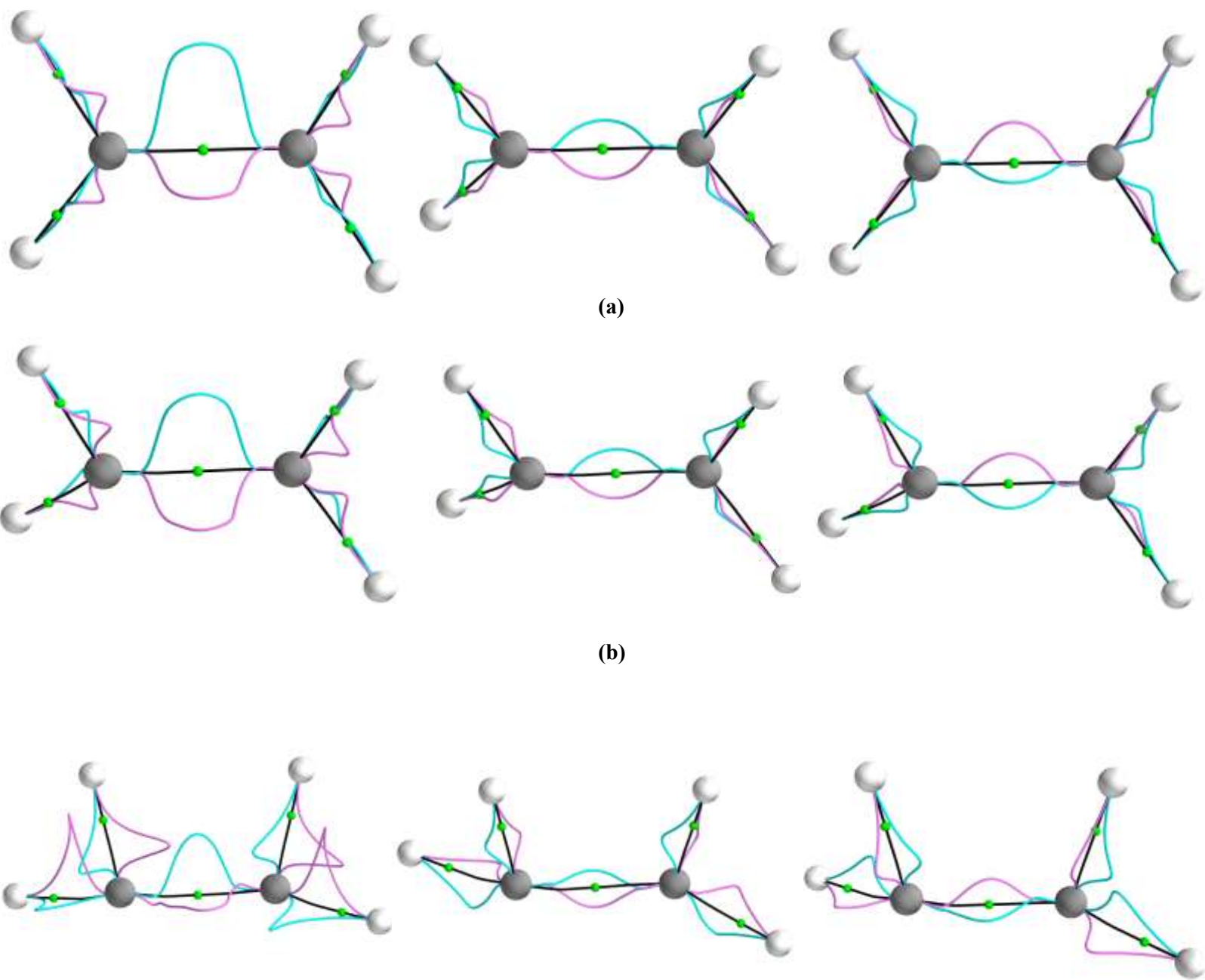

(a)

(b)

(c)

**Figure S5(a).** Magnified (x5) versions of the *p*- (pale-blue) and *q*-paths (magenta) along the bond-path (*r*) corresponding to the *BCPs* of the ethene in clockwise(CW) directions at the value of the torsion θ = 30.0°, 60.0°, 150.0° are presented in sub-figures **(a)-(b)** respectively. The plots are ordered (*p*, *q*), (*p*$_\sigma$, *q*$_\sigma$) and (*p*$_{\sigma H}$, *q*$_{\sigma H}$) respectively.

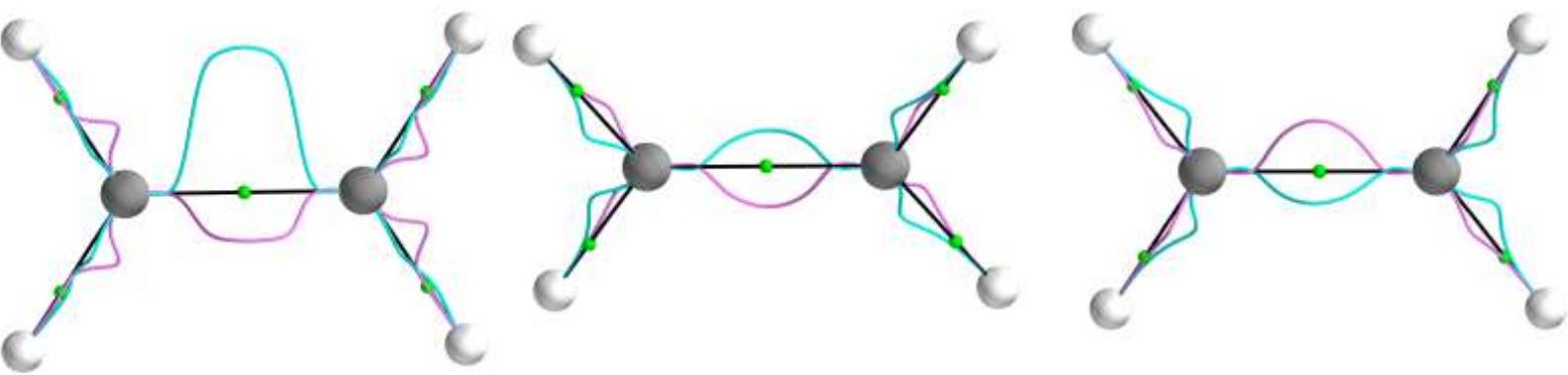

(a)

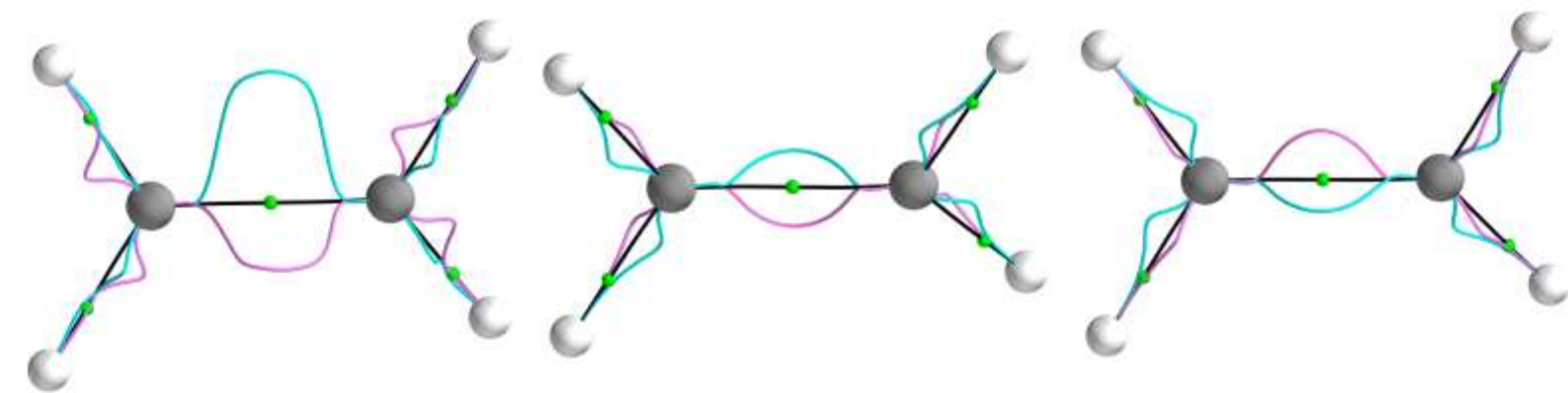

(b)

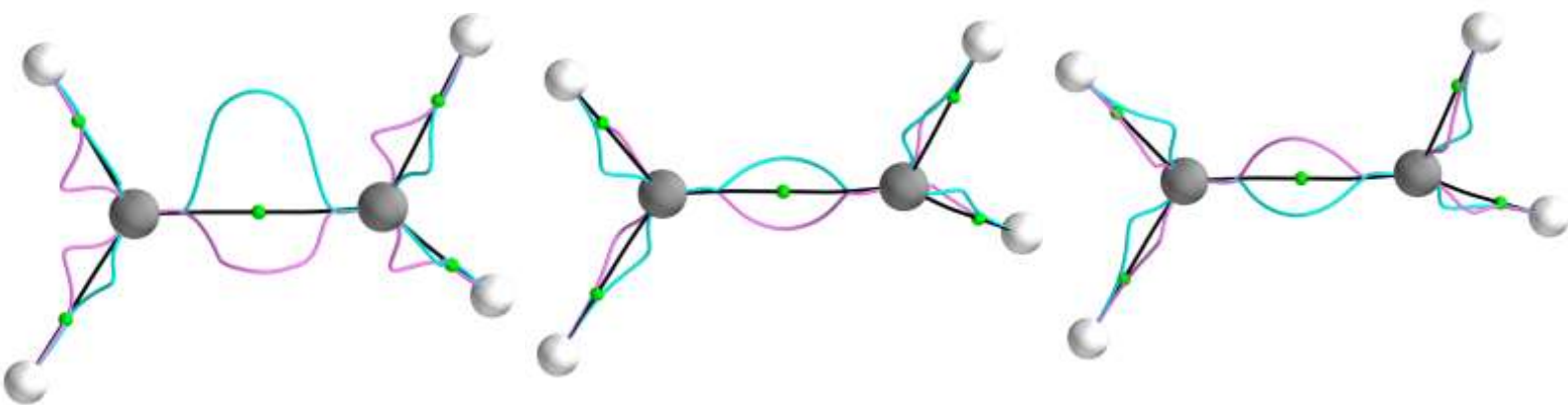

(c)

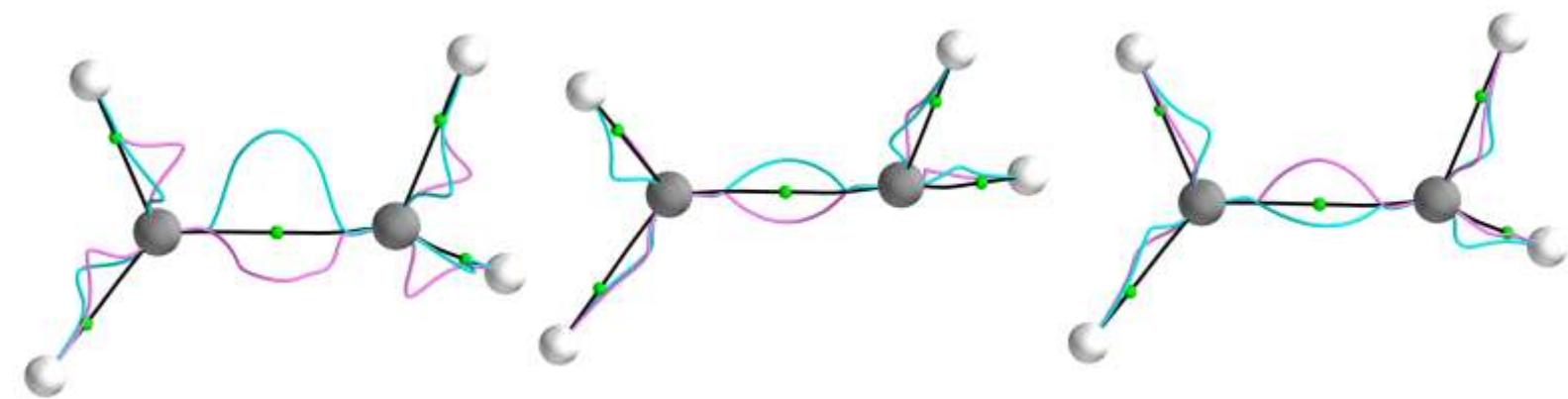

(d)

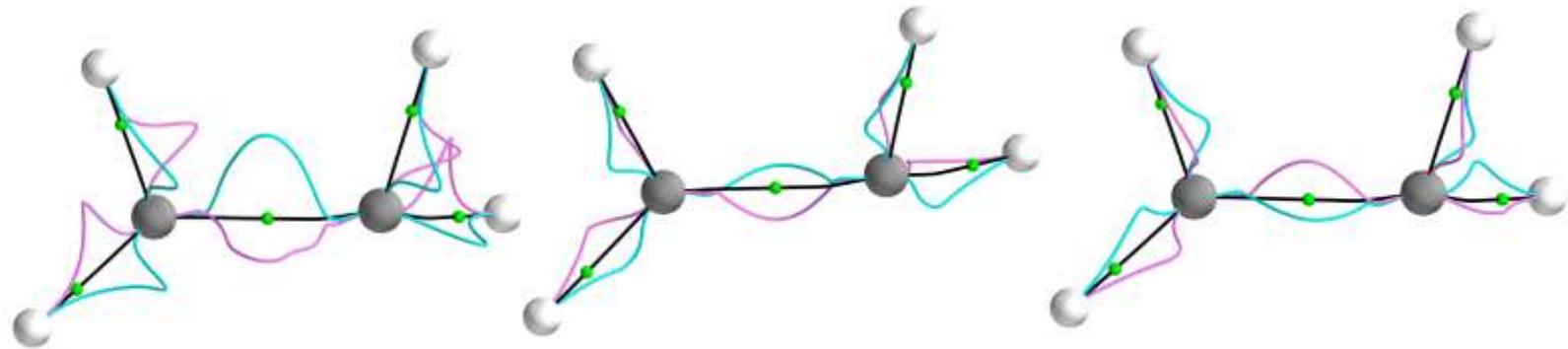

**(e)**

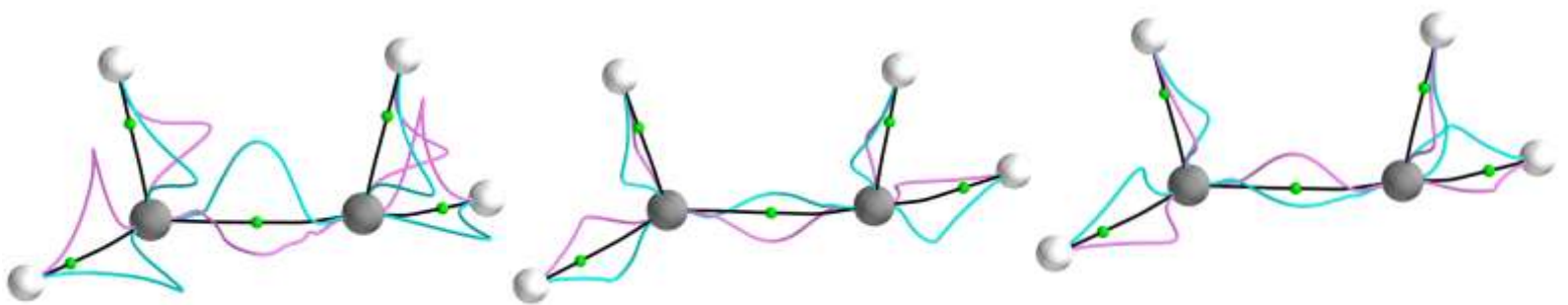

**(f)**

**Figure S5(b).** Magnified (x5) versions of the *p*- (pale-blue) and *q*-paths (magenta) along the bond-path (*r*) corresponding to the *BCPs* of the ethene in counterclockwise directions (CCW) directions at the value of the torsion θ = 0.0°, 30.0°, 60.0°, 90.0°, 120.0°, 150.0° are presented in sub-figures **(a)-(f)** respectively. The plots are ordered ($p$, $q$), ($p_\sigma$, $q_\sigma$) and ($p_{\sigma H}$, $q_{\sigma H}$) respectively.

**6. Supplementary Materials S6.**

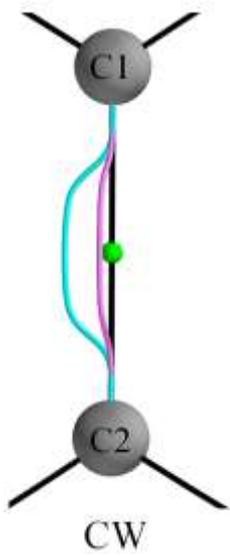 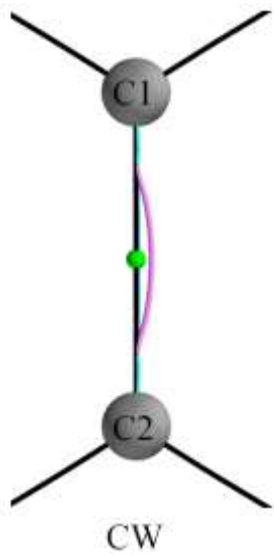 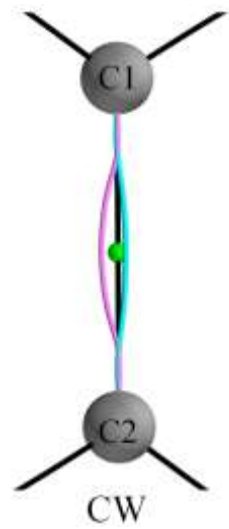

(a)

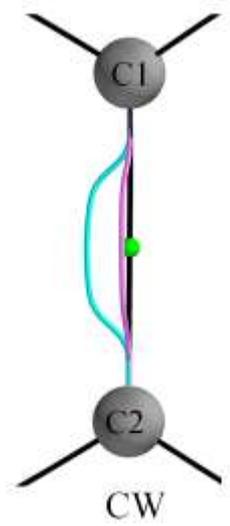 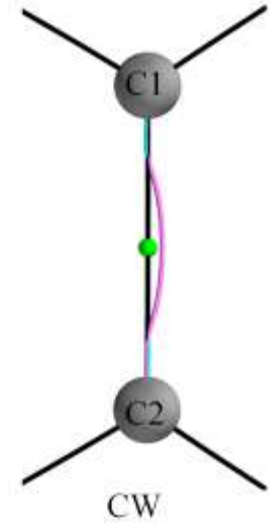 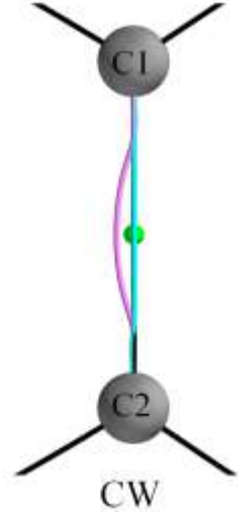

(b)

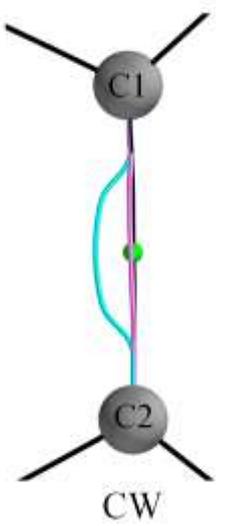 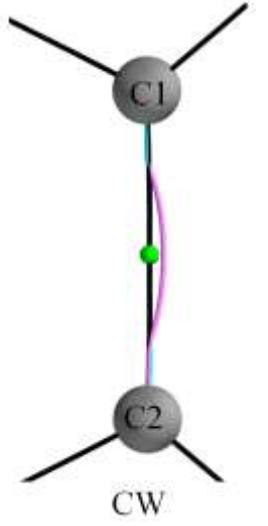 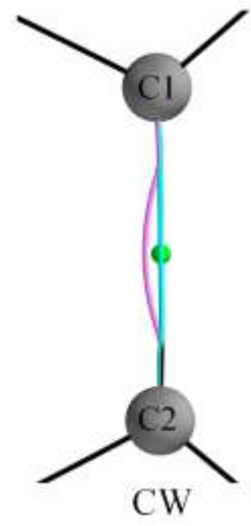

(c)

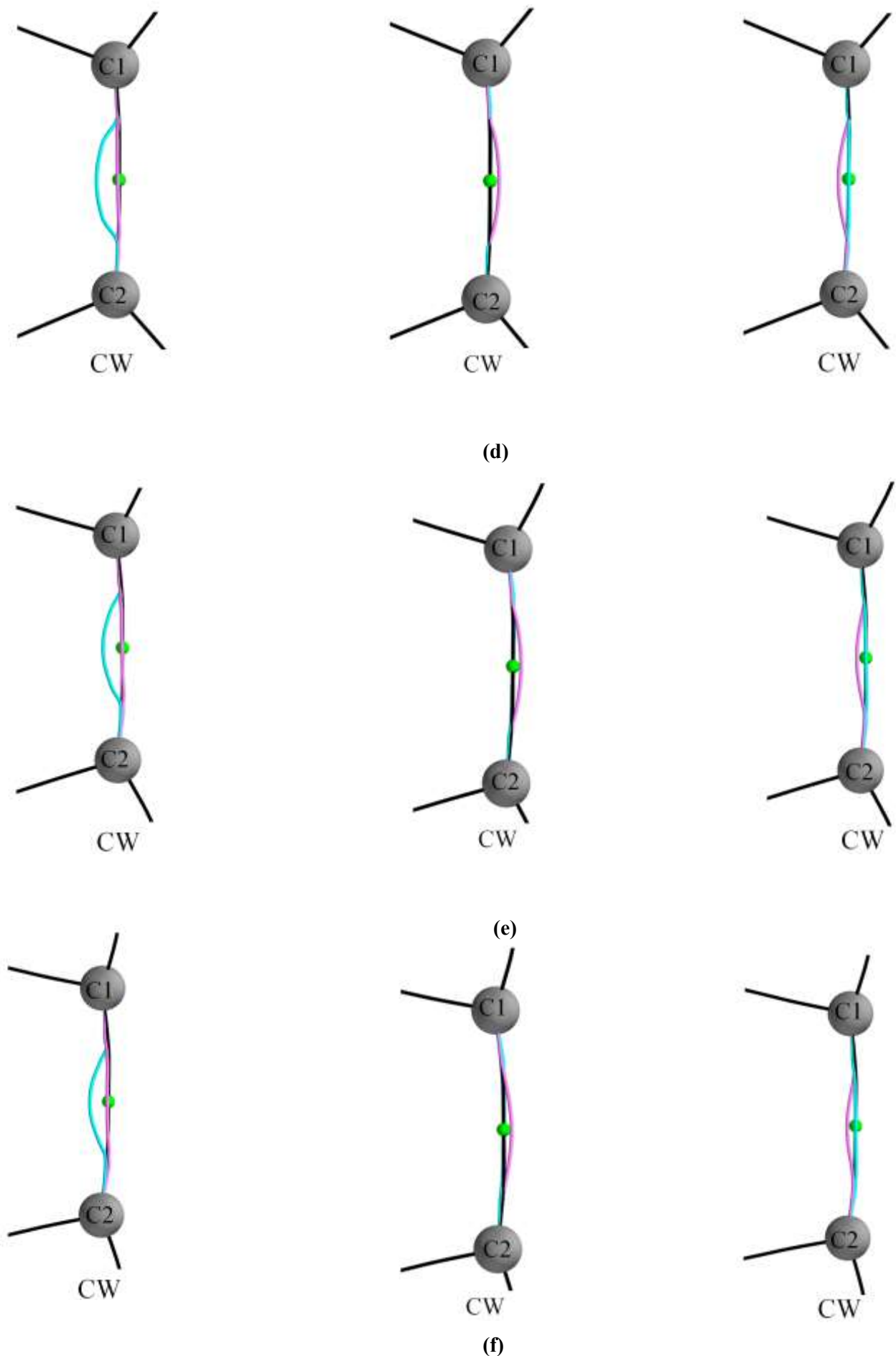

**Figure S6(a).** The *p*- (pale-blue) and *q*-paths (magenta) along the bond-path (*r*) corresponding to the *BCPs* for the C1-C2 *BCP* of the ethene in clockwise(CW) directions at the value of the torsion θ = 0.0°, 30.0°, 60.0°, 90.0°, 120.0°,150.0° are presented in sub-figures **(a)-(f)** respectively. The plots are ordered (*p*, *q*), (*p*$_\sigma$, *q*$_\sigma$) and (*p*$_{\sigma H}$, *q*$_{\sigma H}$) respectively.

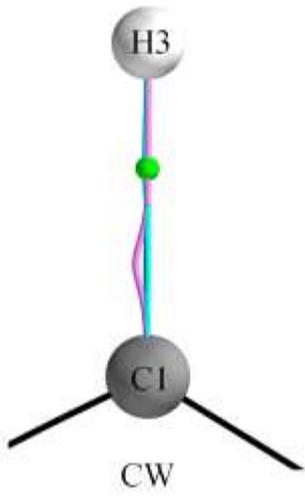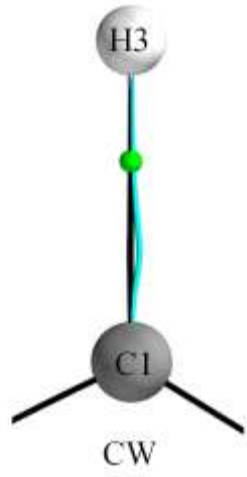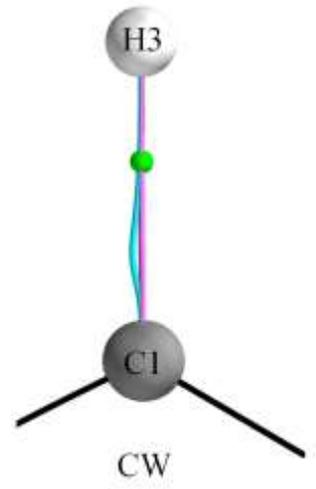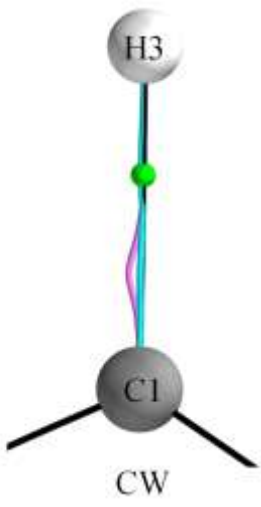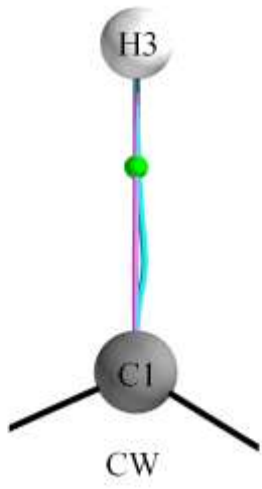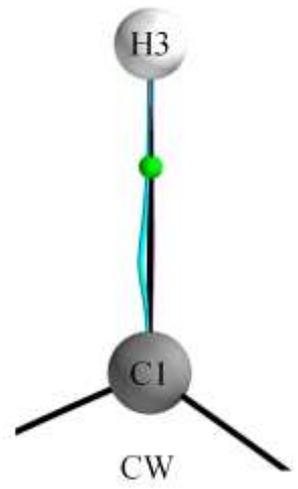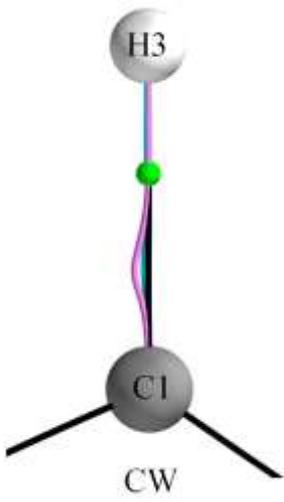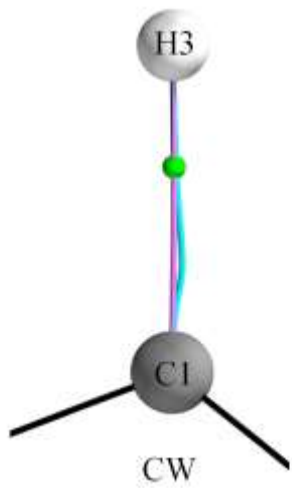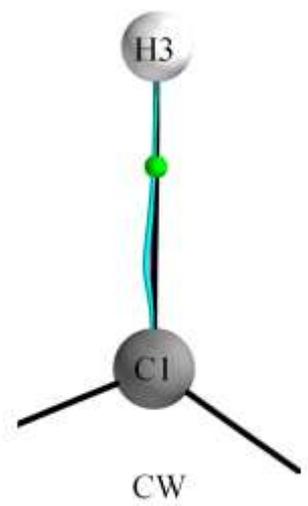

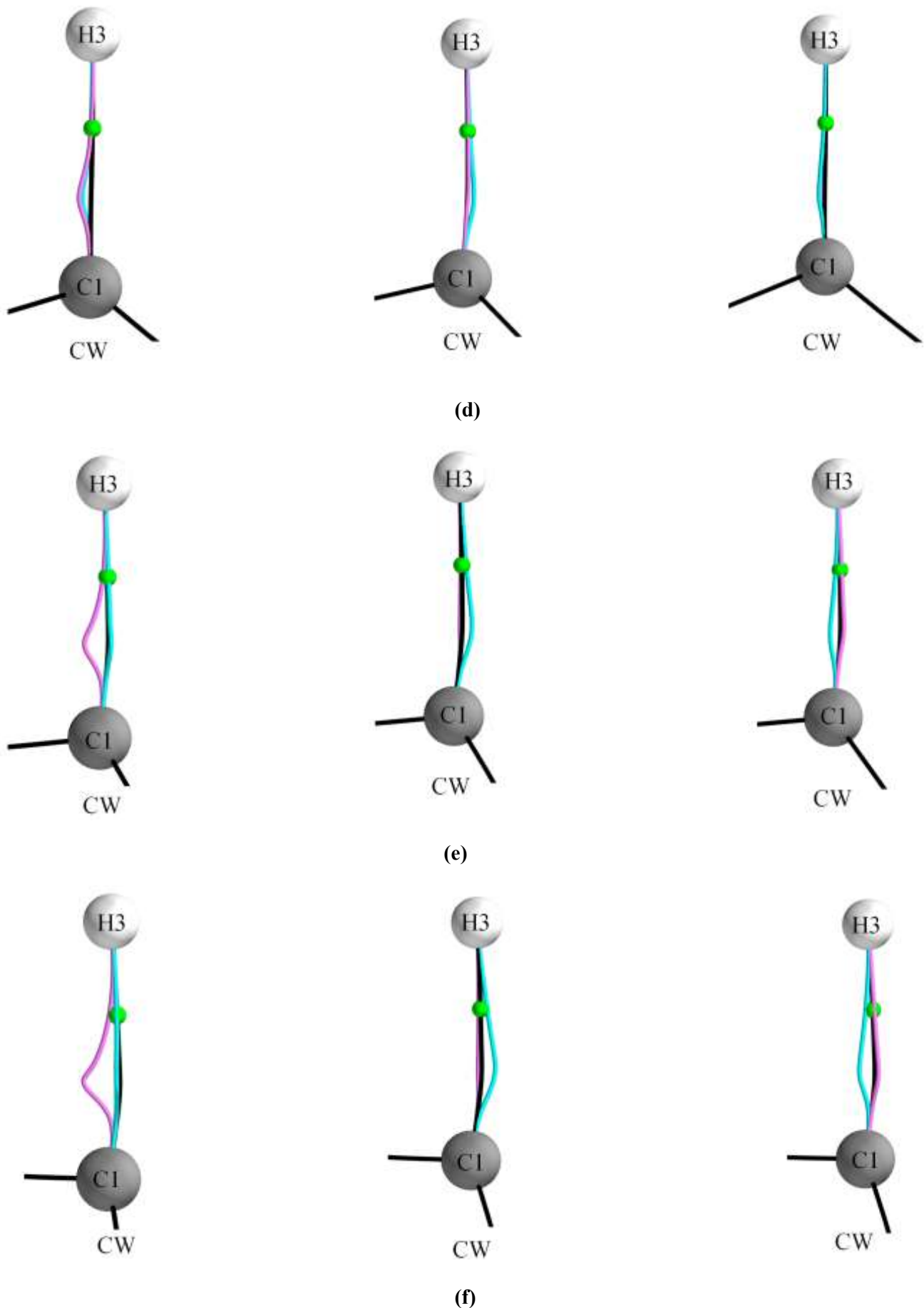

**(d)**

**(e)**

**(f)**

**Figure S6(b).** The *p*- (pale-blue) and *q*-paths (magenta) along the bond-path (*r*) corresponding to the *BCPs* for the C1-H3 *BCP* of the ethene in clockwise(CW) directions at the value of the torsion θ = 0.0°, 30.0°, 60.0°, 90.0°, 120.0°,150.0° are presented in sub-figures **(a)-(f)** respectively. The plots are ordered (*p*, *q*), ($p_\sigma$, $q_\sigma$) and ($p_{\sigma H}$, $q_{\sigma H}$) respectively.

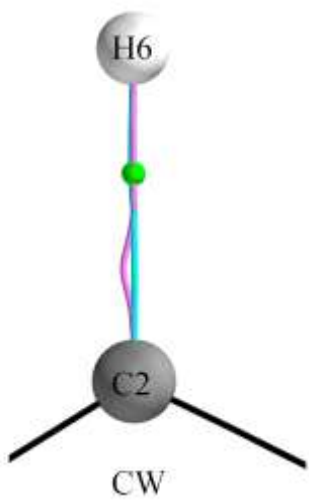 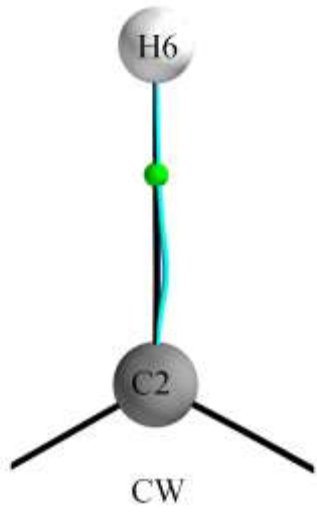 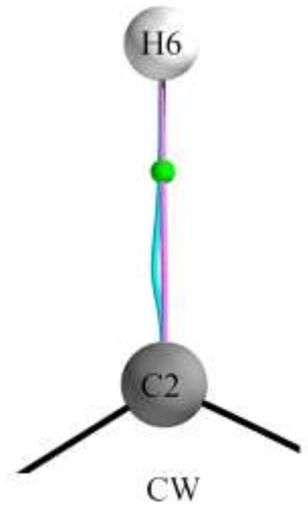

(a)

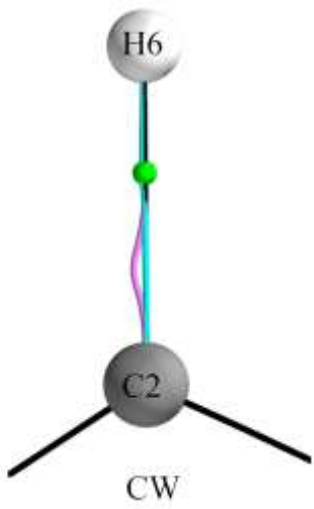 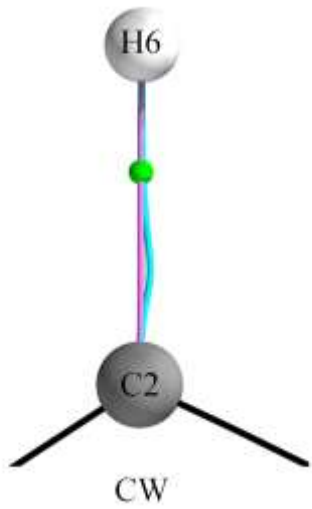 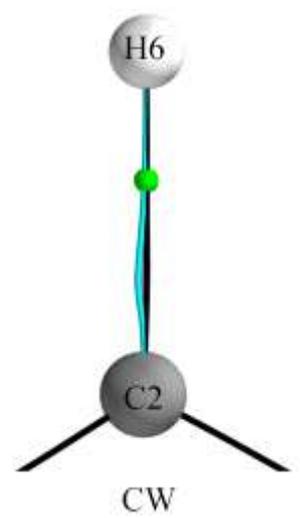

(b)

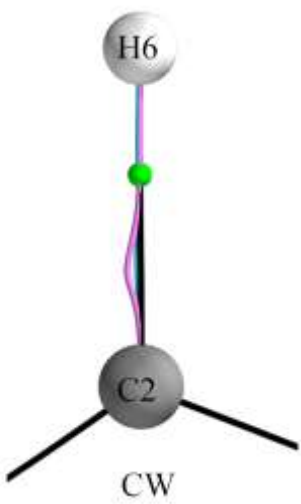 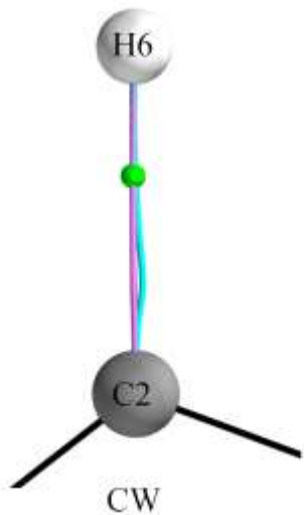 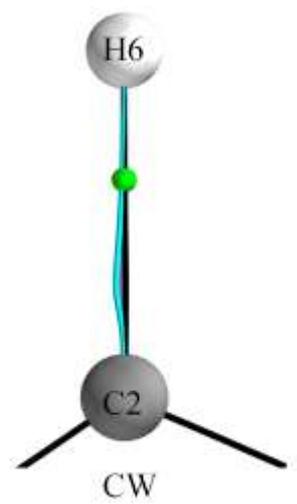

(c)

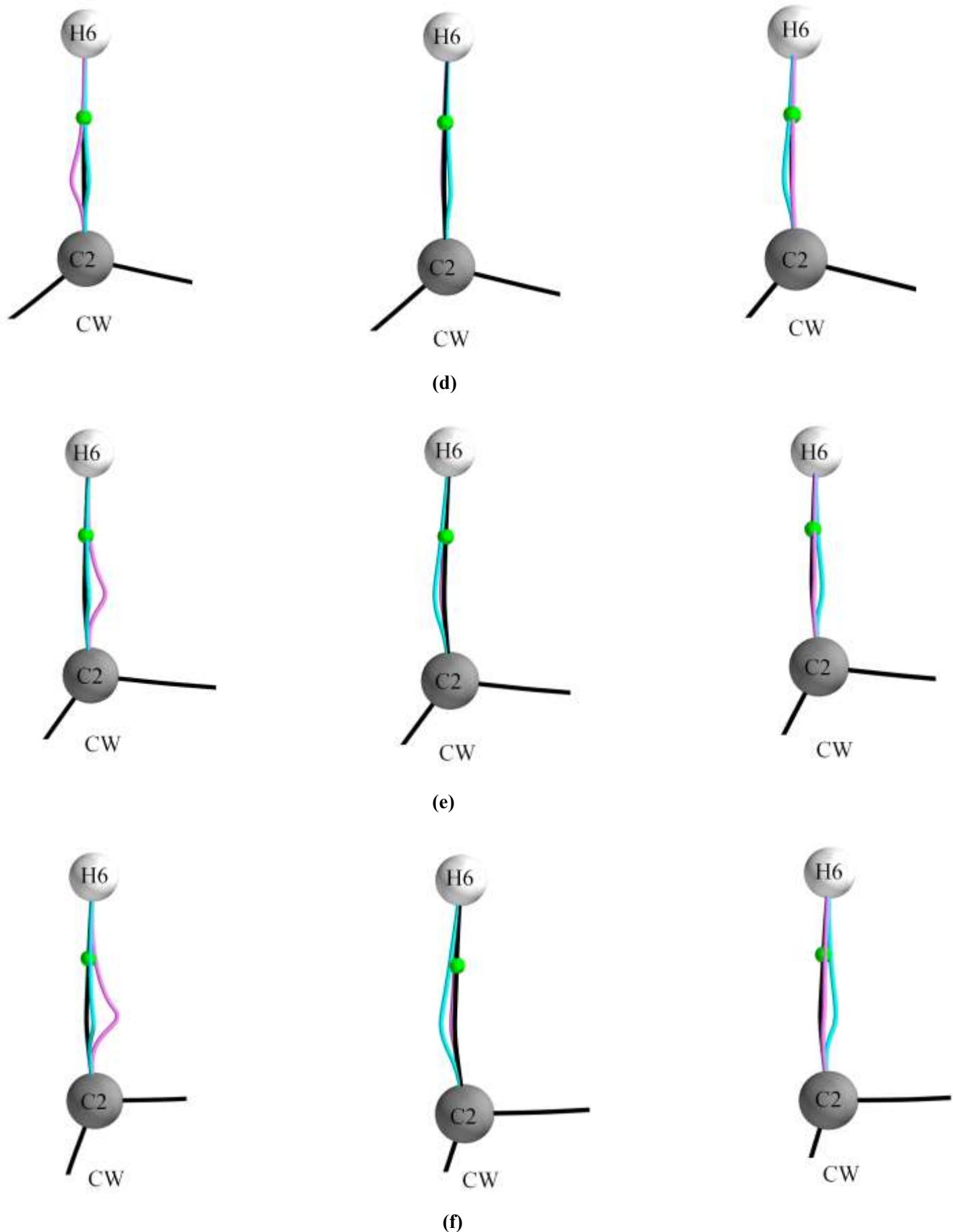

**Figure S6(c).** The *p*- (pale-blue) and *q*-paths (magenta) along the bond-path (*r*) corresponding to the *BCPs* for the C2-H6 *BCP* of the ethene in clockwise(CW) directions at the value of the torsion θ = 0.0°, 30.0°, 60.0°, 90.0°, 120.0°,150.0° are presented in sub-figures **(a)-(f)** respectively. The plots are ordered (*p*, *q*), (*p*$_\sigma$, *q*$_\sigma$) and (*p*$_{\sigma H}$, *q*$_{\sigma H}$) respectively.

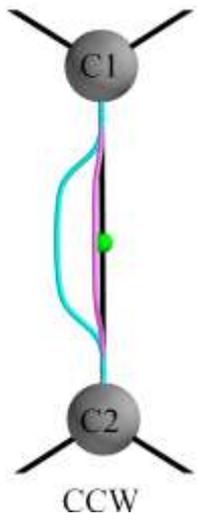 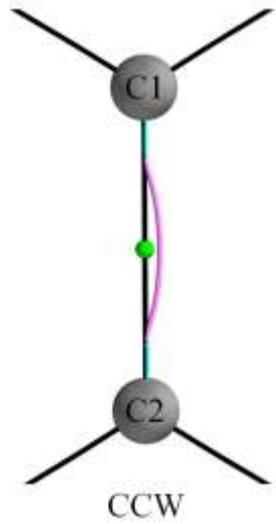 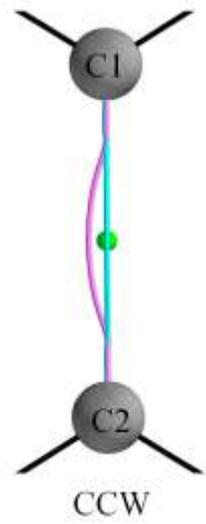

**(a)**

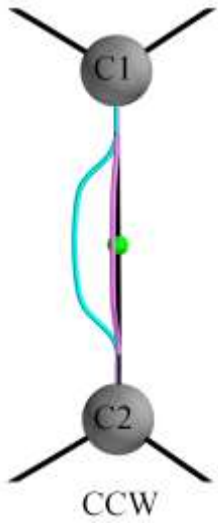 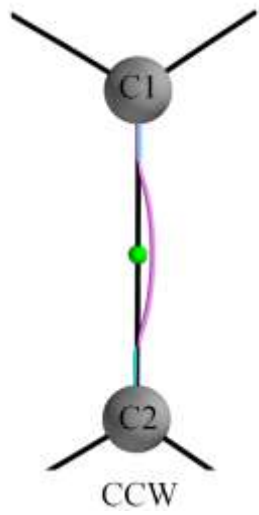 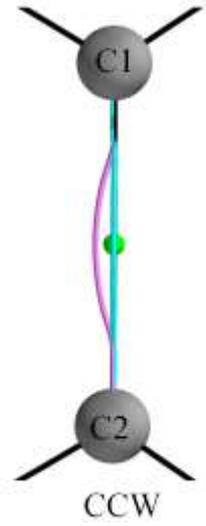

**(b)**

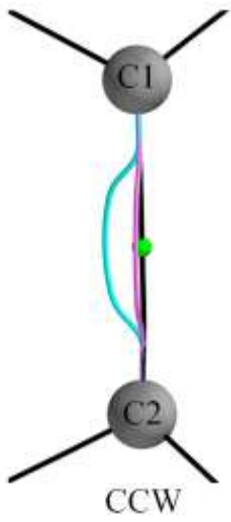 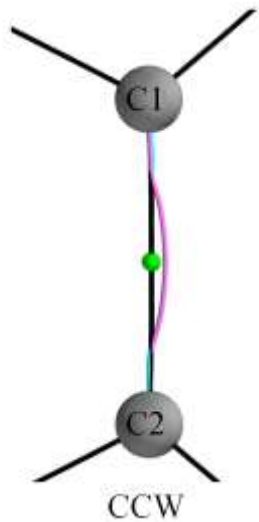 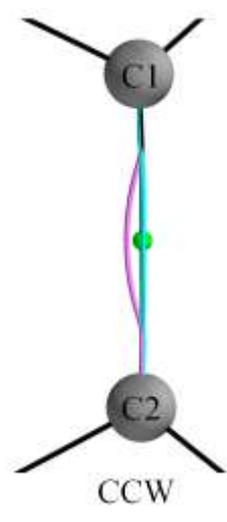

**(c)**

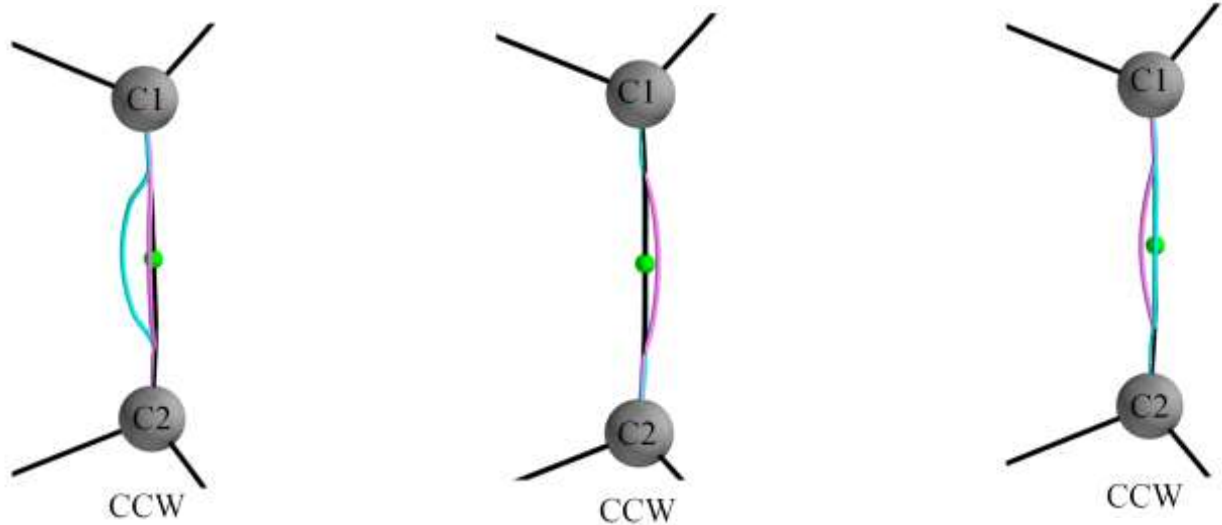

**(d)**

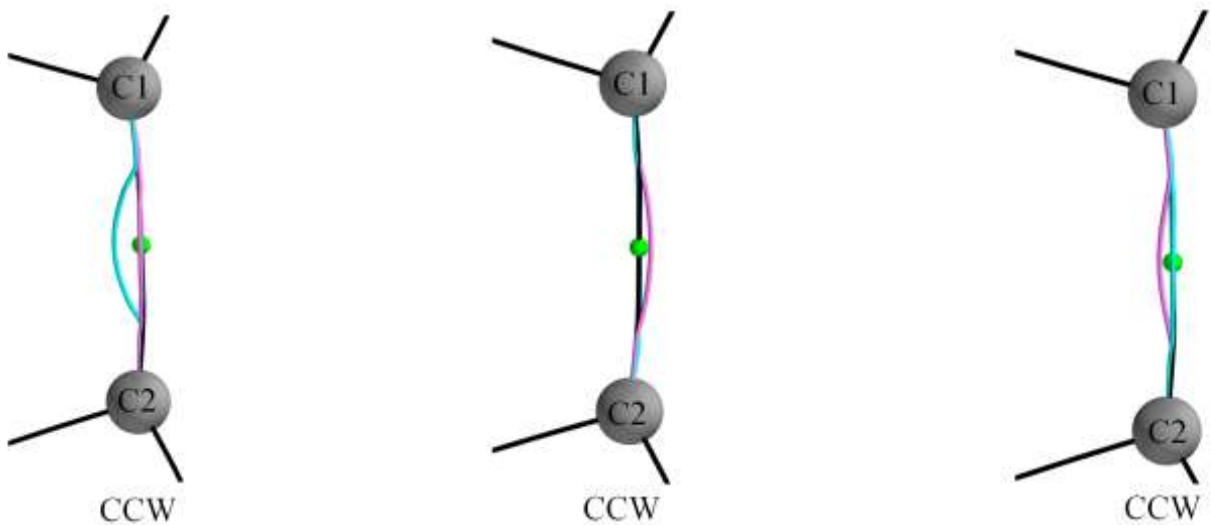

**(e)**

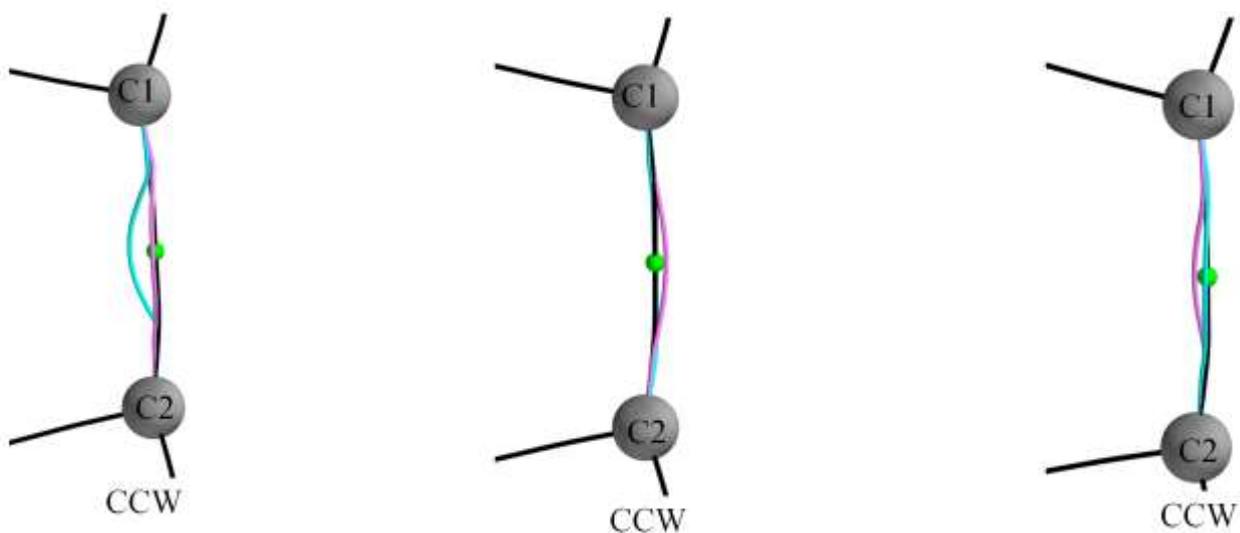

**(f)**

**Figure S6(d).** The $p$- (pale-blue) and $q$-paths (magenta) along the bond-path ($r$) corresponding to the *BCPs* for the C1-C2 *BCP* of the ethene in counterclockwise directions (CCW) directions at the value of the torsion θ = 0.0°, 30.0°, 60.0°, 90.0°, 120.0°, 150.0° are presented in sub-figures **(a)-(f)** respectively. The plots are ordered ($p$, $q$), ($p_\sigma$, $q_\sigma$) and ($p_{\sigma H}$, $q_{\sigma H}$) respectively.

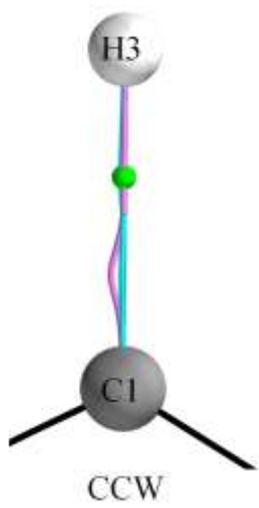 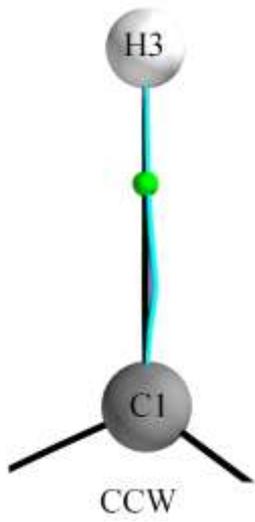 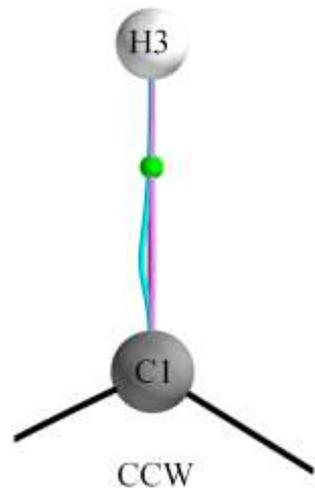

**(a)**

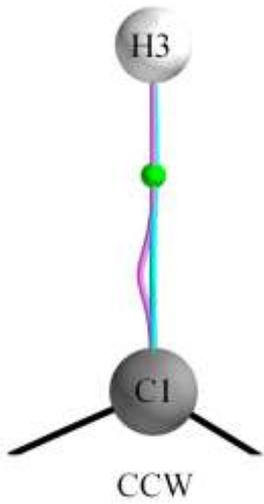 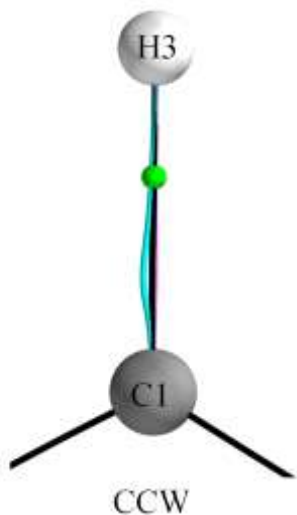 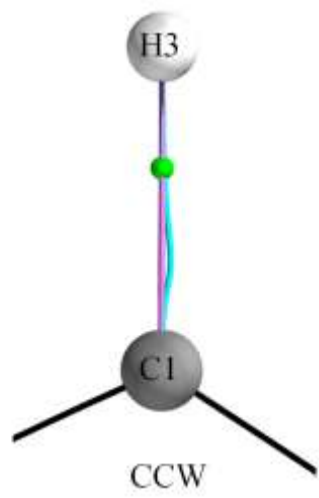

**(b)**

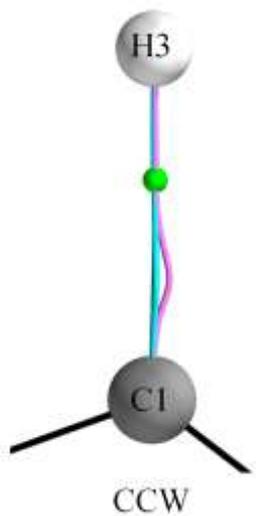 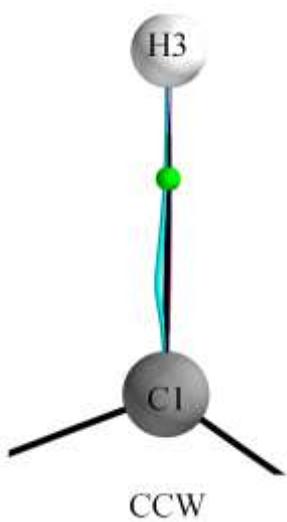 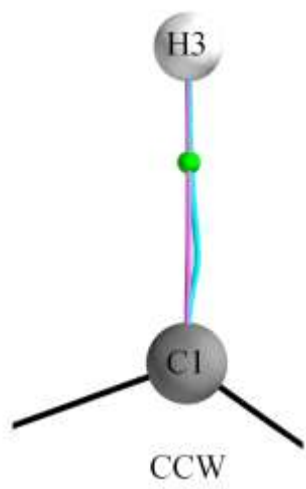

**(c)**

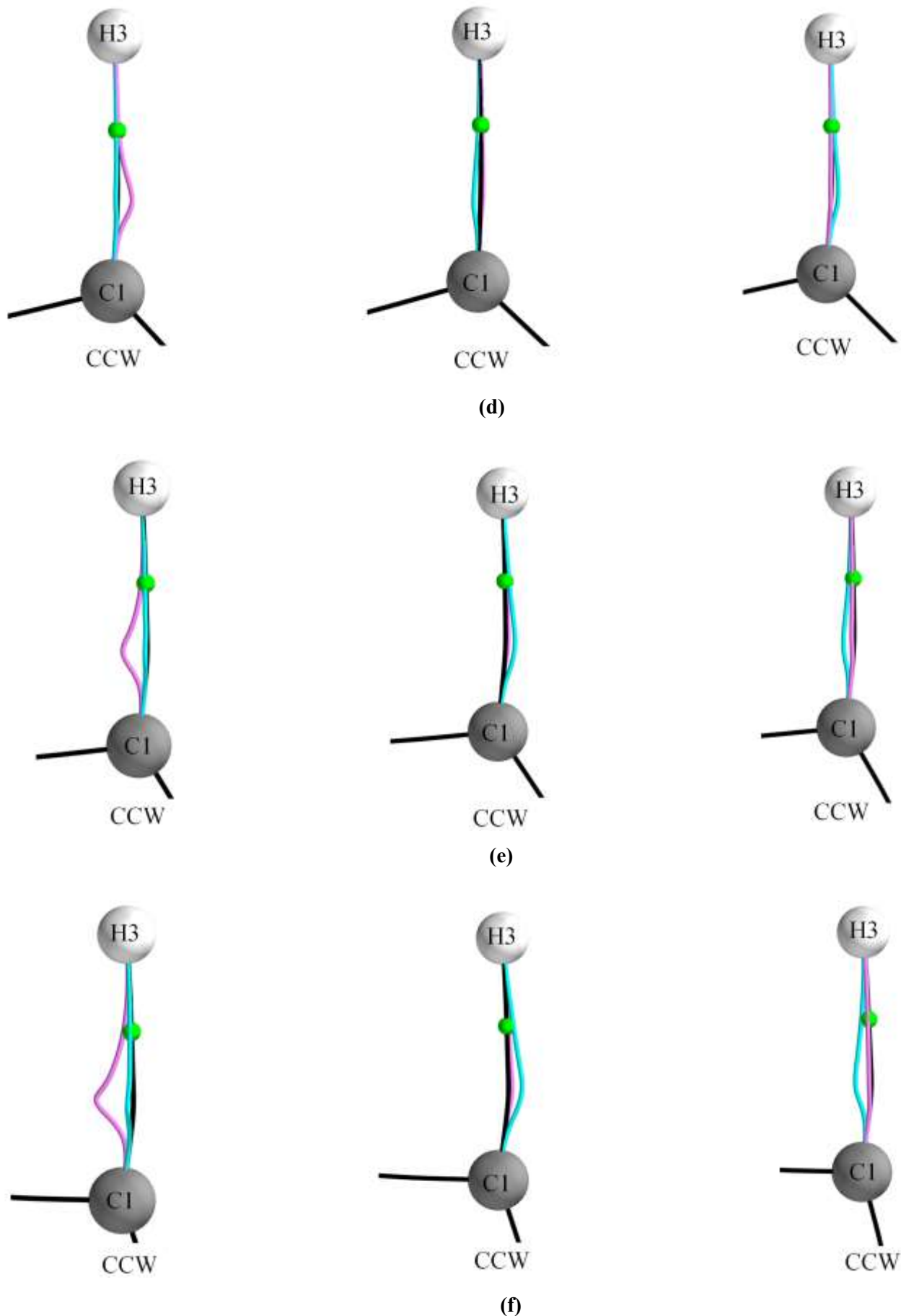

**Figure S6(e)** The *p*- (pale-blue) and *q*-paths (magenta) along the bond-path (*r*) corresponding to the *BCPs* for the C1-H3 *BCP* of the ethene in counterclockwise directions (CCW) directions at the value of the torsion θ = 0.0°, 30.0°,60.0°, 90.0°, 120.0°,150.0° are presented in sub-figures **(a)-(f)** respectively. The plots are ordered (*p*, *q*), (*p*$_\sigma$, *q*$_\sigma$) and (*p*$_{\sigma H}$, *q*$_{\sigma H}$) respectively.

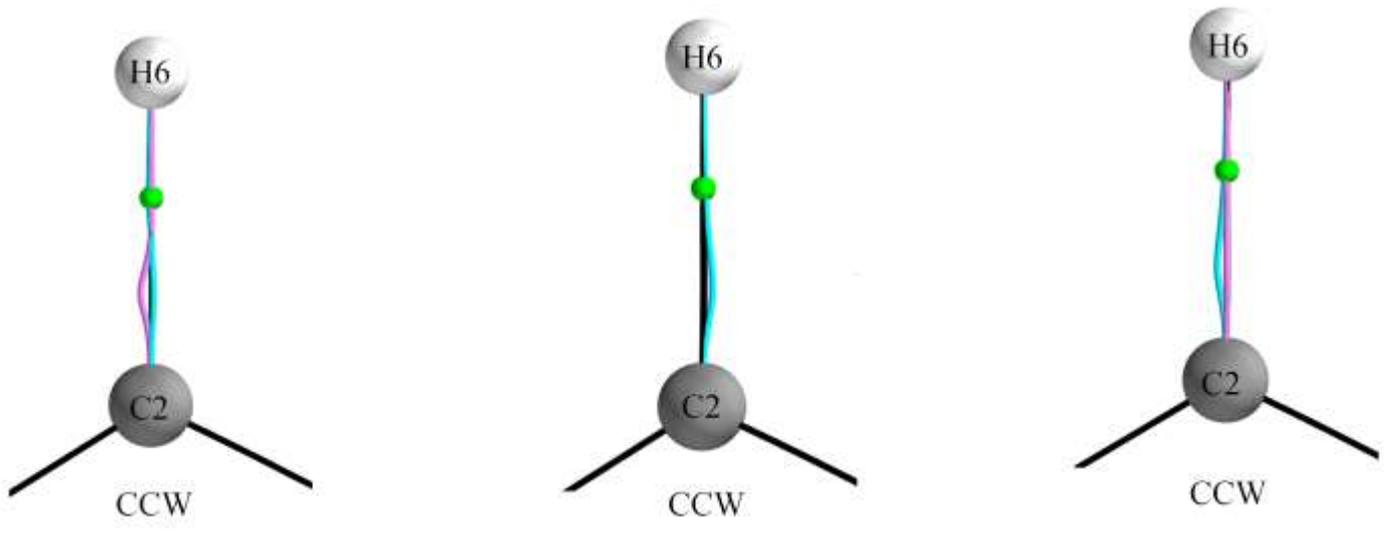

(a)

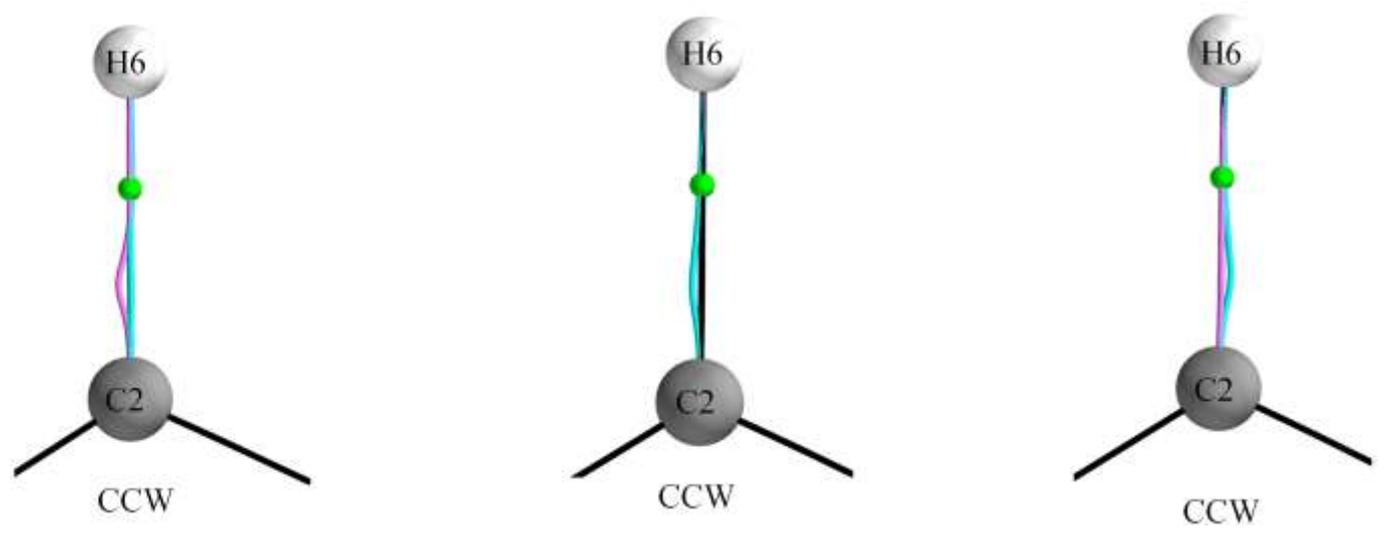

(b)

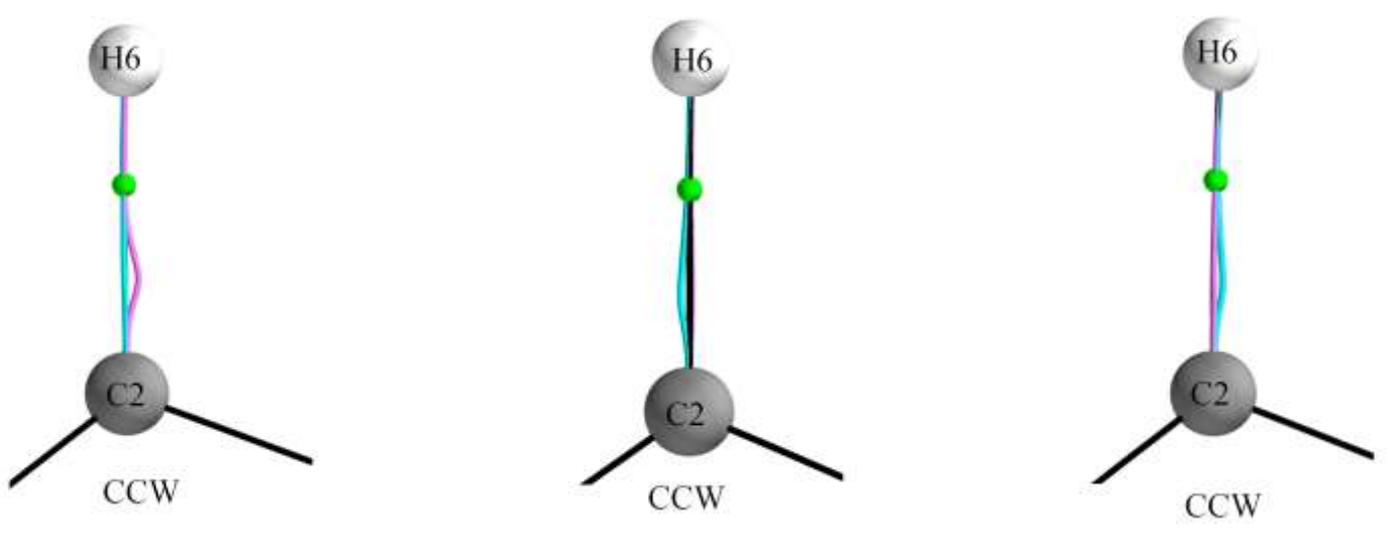

(c)

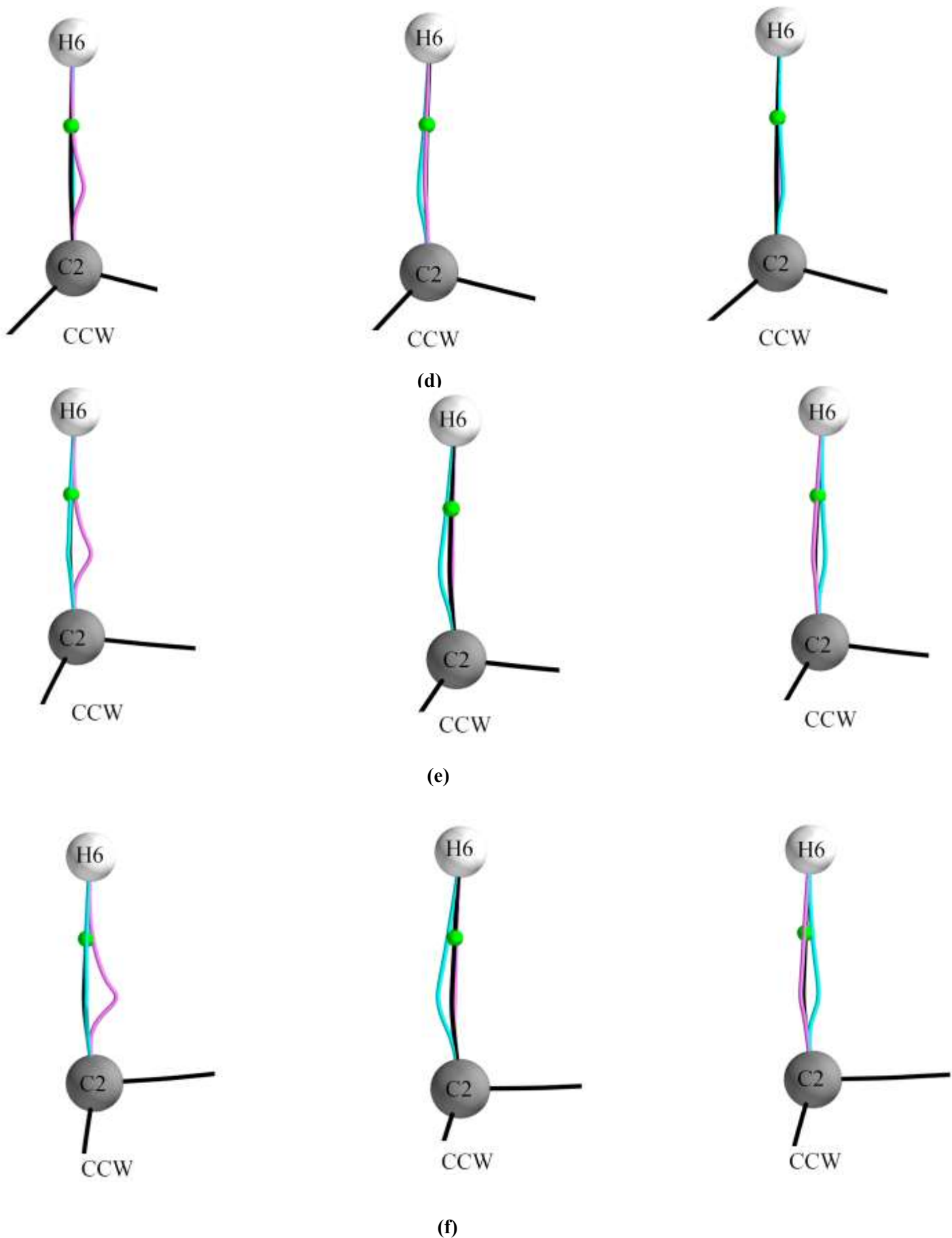

**Figure S6(f).** The *p*- (pale-blue) and *q*-paths (magenta) along the bond-path (*r*) corresponding to the *BCPs* for the C2-H6 *BCP* of the ethene in counterclockwise directions (CCW) directions at the value of the torsion θ = 0.0°, 30.0°, 60.0°, 90.0°, 120.0°,150.0° are presented in sub-figures **(a)-(f)** respectively. The plots are ordered (*p*, *q*), ($p_\sigma$, $q_\sigma$) and ($p_{\sigma H}$, $q_{\sigma H}$) respectively.

**7. Supplementary Materials7.**

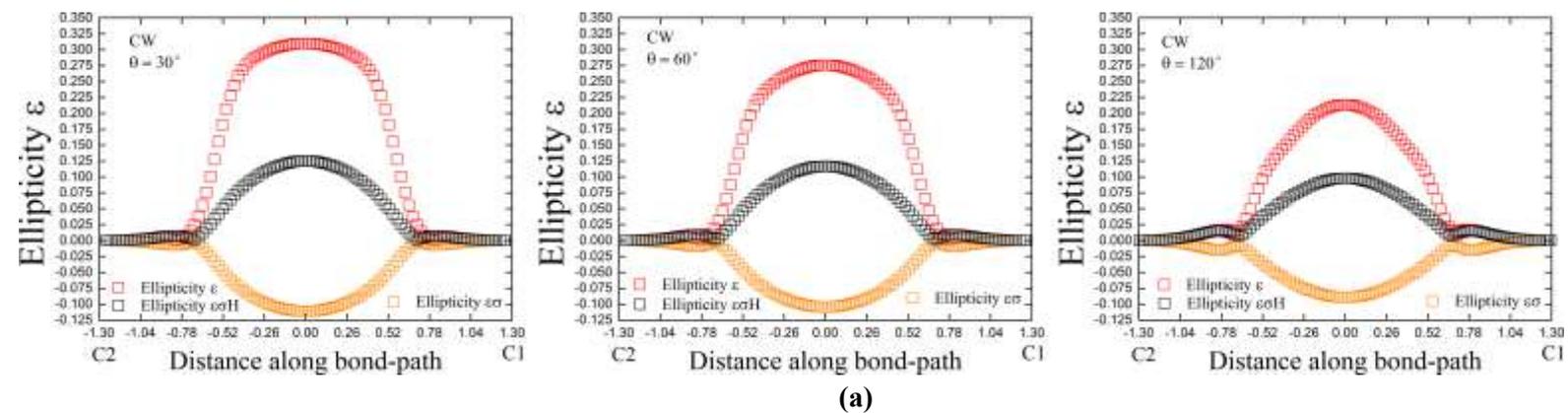

**(a)**

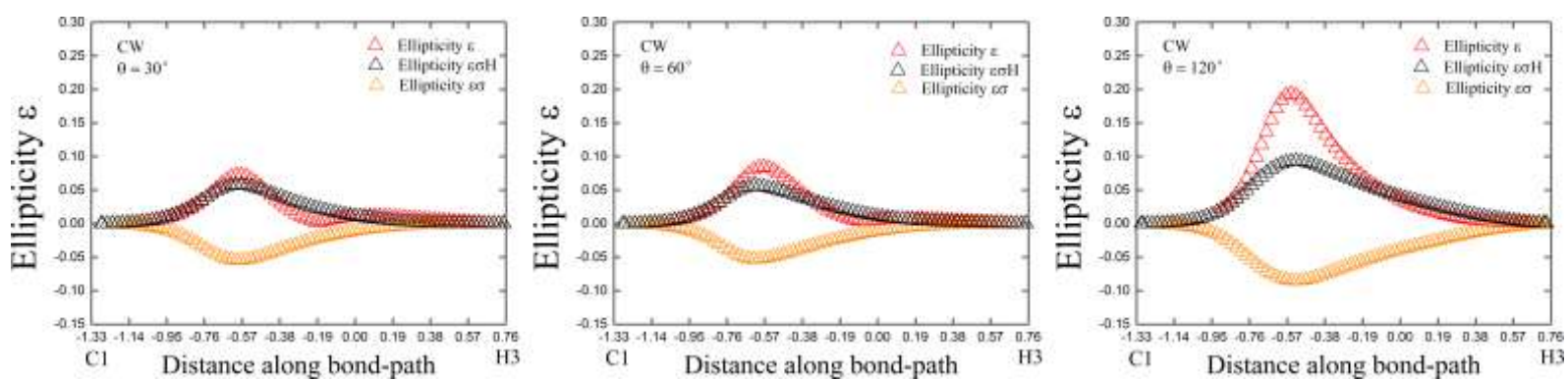

**(b)**

**Figure S7(a).** The variations of the three ellipticity ε profiles for the ethene in clockwise(CW) directions along the bond-path (*r*) associated with the C1-C2 *BCP* and C1-H3 *BCP*, where θ = 30.0°, 60.0° and 120.0° are presented in sub-figures **(a)-(c)** respectively.

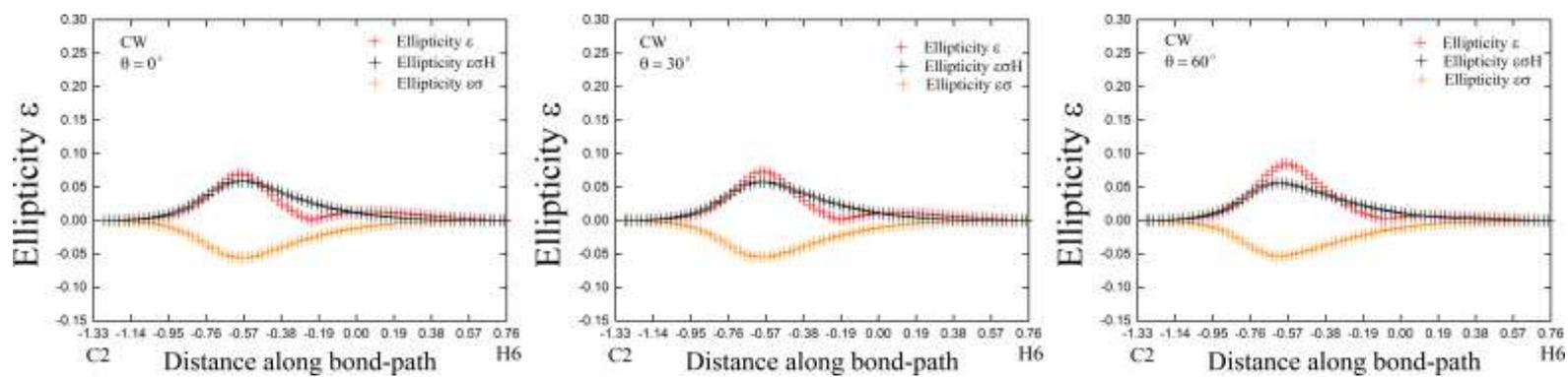

(a)

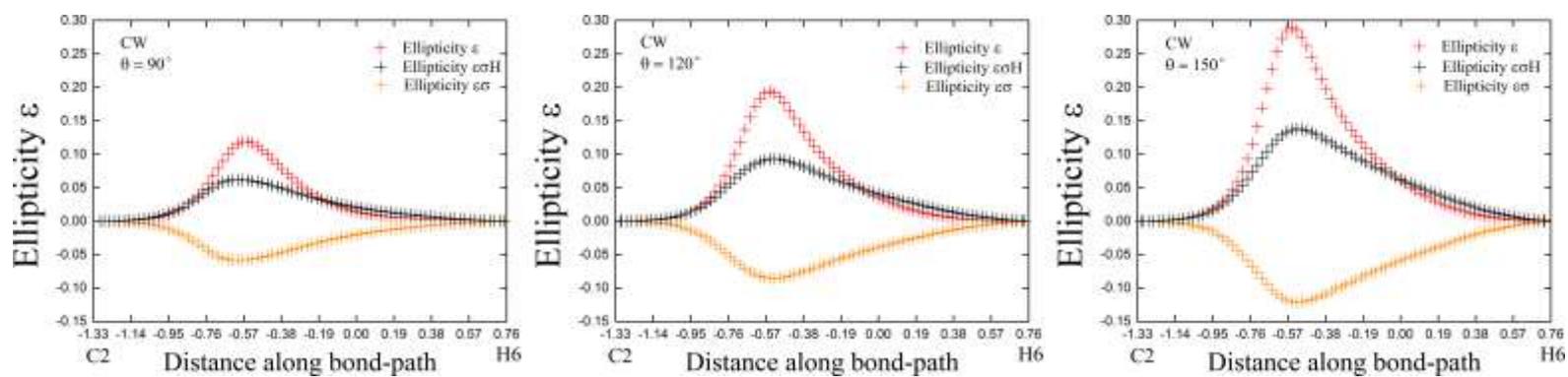

(b)

**Figure S7(b).** The variations of the three ellipticity ε profiles for the ethene in clockwise(CW) directions along the bond-path (*r*) associated with the C2-H6 *BCP*, where θ = 0.0°, 30.0°, 60.0°, 90.0°, 120.0° and 150.0° are presented in sub-figures **(a)-(b)** respectively.

## 8. Supplementary Materials S8. Implementation details of the calculation of the eigenvector-following path lengths $\mathbb{H}$ and $\mathbb{H}^*$.

When the QTAIM eigenvectors of the Hessian of the charge density $\rho(\mathbf{r})$ are evaluated at points along the bond-path, this is done by requesting them via a spawned process which runs the selected underlying QTAIM code, which then passes the results back to the analysis code. For some datasets, it occurs that, as this evaluation considers one point after another in sequence along the bond-path, the returned calculated $\underline{\mathbf{e}_2}$ (correspondingly $\underline{\mathbf{e}_1}$ is used to obtain $\mathbb{H}^*$) eigenvectors can experience a 180-degree 'flip' at the 'current' bond-path point compared with those evaluated at both the 'previous' and 'next' bond-path points in the sequence. These 'flipped' $\underline{\mathbf{e}_2}$ (or $\underline{\mathbf{e}_1}$) eigenvectors, caused by the underlying details of the numerical implementation in the code that computed them, are perfectly valid, as these are defined to within a scale factor of -1 (i.e. inversion). The analysis code used in this work detects and re-inverts such temporary 'flips' in the $\underline{\mathbf{e}_2}$ (or $\underline{\mathbf{e}_1}$) eigenvectors to maintain consistency with the calculated $\underline{\mathbf{e}_2}$ (or $\underline{\mathbf{e}_1}$) eigenvectors at neighboring bond-path points, in the evaluation of path eigenvector-following path lengths $\mathbb{H}$ and $\mathbb{H}^*$.

The corresponding stress tensor lengths $\mathbb{H}_\sigma$, $\mathbb{H}_{\sigma H}$ and $\mathbb{H}_\sigma^*$, $\mathbb{H}_{\sigma H}^*$ are obtained using the same method.